\documentclass[english,aps,prb,twocolumn,superscriptaddress,longbibliography]{revtex4-1}
\usepackage[T1]{fontenc}
\usepackage[latin9]{inputenc}
\usepackage{times}
\usepackage{babel}
\usepackage{array}
\usepackage{amsmath}
\usepackage{amssymb}
\usepackage{comment}
\usepackage[table,usenames,dvipsnames]{xcolor}

\usepackage{graphicx}
\usepackage[unicode=true,pdfusetitle,
 bookmarks=true,bookmarksnumbered=false,bookmarksopen=false,
 breaklinks=false,pdfborder={0 0 1},backref=false,colorlinks=false]
 {hyperref}

\definecolor{jens}{rgb}{0,.8,.5}

\makeatletter

\providecommand{\tabularnewline}{\\}

\newcommand{\Tr}{\mathrm{Tr}}
\usepackage{braket}
\usepackage[caption=false]{subfig}

\@ifundefined{showcaptionsetup}{}{%
 \PassOptionsToPackage{caption=false}{subfig}}
\usepackage{subfig}
\makeatother

\begin{document}
\global\long\def\k#1{\ket{#1}}%
\global\long\def\b#1{\bra{#1}}%
\global\long\def\bk#1{\braket{#1}}%
\global\long\def\Tr{\mathrm{Tr}}%
\global\long\def\Var{\text{Var}}%

\preprint{This line only printed with preprint option}

\title{Entanglement estimation in tensor network states via sampling}


\author{Noa Feldman}
\affiliation{\textsuperscript{}Raymond and Beverly Sackler School of Physics and
Astronomy, Tel-Aviv University, 6997801 Tel Aviv, Israel}
\author{Augustine Kshetrimayum}
\affiliation{\textsuperscript{}Dahlem Center for Complex Quantum Systems, Freie
Universit{\"a}t Berlin, 14195 Berlin, Germany\textsuperscript{}}
\affiliation{Helmholtz Center Berlin, 14109 Berlin, Germany}
\author{Jens Eisert}
\affiliation{\textsuperscript{}Dahlem Center for Complex Quantum Systems, Freie
Universit{\"a}t Berlin, 14195 Berlin, Germany\textsuperscript{}}
\affiliation{Helmholtz Center Berlin, 14109 Berlin, Germany}
\author{Moshe Goldstein}
\affiliation{\textsuperscript{}Raymond and Beverly Sackler School of Physics and
Astronomy, Tel-Aviv University, 6997801 Tel Aviv, Israel}

\begin{abstract}
We introduce a method for extracting meaningful entanglement measures of tensor network states in general dimensions. Current methods require the explicit reconstruction of the density matrix, which is highly demanding, or the contraction of replicas, which requires an effort exponential in the number of replicas and which is costly in terms of memory. In contrast, our method requires the stochastic sampling of matrix elements of the classically represented reduced states with respect to random states drawn from simple product probability measures constituting frames. Even though not corresponding to physical operations, such matrix elements are straightforward to calculate for tensor network states, and their moments provide the R\'enyi entropies and negativities as well as their symmetry-resolved components. We test our method on the one-dimensional critical XX chain and the two-dimensional toric code in a checkerboard geometry. Although the cost is exponential in the subsystem size, it is sufficiently moderate so that - in contrast with other approaches - accurate results can be obtained on a personal computer for relatively large subsystem sizes.
\end{abstract}
\maketitle

\section{introduction\label{sec:introduction}}

Entanglement is the key feature of quantum mechanics that renders it different from classical theories.
It takes centre stage in quantum information processing where it plays the role of a resource.  
The significance of notions of entanglement for capturing properties of condensed matter systems has also long been noted and appreciated \cite{entanglement_as_resource_1993,Horodecki4_2009}. The observation that ground states of gapped phases of matter are expected to feature little entanglement -- in fact, they feature what are called area laws for entanglement entropies \cite{TN_2010} -- is at the basis of \emph{tensor networks} (TN) methods \cite{TN_2008,Orus-AnnPhys-2014} accurately describing interacting quantum many-body systems. It has been noted that certain scalings of entanglement measures can indicate the presence of quantum phase transitions \cite{entanglement_for_physics_2002_sep,mbentanglement_2008}.
Indeed,
the very fact that locally interacting quantum many body systems tend to be much less entangled than they could 
possibly be
renders TN methods a powerful technique to capture their properties \citep{TN_2008,TN_2010,TN_2013,Orus-AnnPhys-2014}.
Maybe most prominently, topologically ordered systems can be regarded as long ranged entangled systems
\cite{WenBook}. In addition, detailed information about the scaling of entanglement properties can provide substantial diagnostic information about properties of condensed matter systems.

Accepting that tensor network states often provide an accurate and efficient classical description of interacting quantum systems, the question arises how one can meaningfully read off \emph{entanglement properties} from tensor network states. This, however, constitutes a challenge. Current entanglement calculation methods in tensor network states in two and higher dimensions are highly impractical even for moderate-size systems, since they
require a full reconstruction of the quantum states at heavy computational costs. 
For R\'{e}nyi entropies one may instead employ the replica trick, which uses several copies of the RDM (as explained in Section ~\ref{subsec:TN} below); this, however, comes with an exponential scaling of the computational effort in the number of copies, often making the calculation unfeasible.


In this work, we develop a method for estimating the entanglement moments of general states represented by tensor networks. We do so by bringing together ideas of tensor network methods with those of
\emph{random measurements} 
\citep{pastMeas_2010,pastMeas_2013,pastMeas_2012_entropy,pastMeas_2015_entropy,
pastMeas_feb_2018_entropy_spin,
pastMeas_mar_2019,pastMeas_apr_2019_experimental_purity,
pastMeas_jun_2019_scrambling,pastMeas_jan_2020_fidelity,
pastMeas9_may_2020_topo,
pastMeas_jun_2020,pastMeas_jun_2020_scrambling,
pastMeas_jul_2020_entropy_experiment}
and \emph{shadow estimation}
\cite{pastMeas_oct_2020_shadows,pastMeas_2013,negativity,GateSets}. In this 
context, it has been understood that entanglement features can be reliably estimated from expectation values of suitable random measurements. 

Here, we bring these ideas to a new level by applying them to quantum states that are classically represented in the first place by tensor networks.
The core idea of these methods
is basically the following: While the entanglement moments naively require several
copies of the system, we can refrain from this requirement by resorting to random sampling. The general protocol is to evolve the system under
a random unitary drawn from the Haar measure followed by a measurement
of a suitable projector. The process is repeated and moments of the
results are averaged over different unitaries, giving as a result
entanglement moments or other density-matrix-based properties. The
effectiveness of this mindset
has been demonstrated experimentally 
in a number of platforms, including
that of cold atoms for R\'enyi
entropies\citep{pastMeas_apr_2019_experimental_purity,pastMeas_jul_2020_entropy_experiment}
and R\'enyi negativities\citep{negativity,negativity21}. 

While these ideas have been further developed into estimation techniques
\cite{pastMeas_oct_2020_shadows} giving rise to classical representations in their own right, we turn these ideas upside down by applying them to quantum systems that are already classically represented by tensor networks. There are 
some crucial differences that arise in classical representations compared to quantum experiments:
First, they are much more suitable for a direct calculation of expectation values, rather than
estimating them from sampling from 
measurements. Second, and importantly, when performing a classical simulation, we are not limited
to physically-allowed actions, and specifically, we are not constrained to the application of unitary operators and measurements. This feature is to an extent reminiscent of shadow estimation in that also there, unphysical
maps are made use of. It is the estimation procedure itself that is not physical here, however.
The method
we develop allows for having only a single copy of the simulated state, and at the same time for estimating the entanglement moments based on matrix elements that are naturally calculated. Instead of sampling operators 
from the Haar measure or some unitary $n$-design \cite{ndesigns}, we only need to
sample from a simple, finite set of tensor products of independent
$d$-dimensional vectors -- specifically from what are called 
\emph{frames} or \emph{spherical complex 1-designs} \cite{frames}, 
where $d$ is the Hilbert space of a single
site in the system. Furthermore, this simple structure allows our method to be applied to arbitrary system geometries.

The remainder of this work is organized as follows. Section \ref{sec:Preliminaries}
includes preliminary theoretical background. The R\'enyi moments
we aim to estimate are defined and their relation to standard entanglement
measures is discussed in Section  \ref{subsec:Entanglement-measures}.
Section  \ref{subsec:TN} covers the basic ideas of the TN ansatzes we
use in our work: For one-dimensional systems, the \emph{matrix product state}
(MPS) ansatz, and in higher dimensions, \emph{projected-entangled-paired-states}
(PEPS) and its infinite system size version known as iPEPS. We discuss the algorithms
we used for extracting the reduced density matrix and the naive method
for estimating entanglement moments of states represented by these
ansatzes. The solvable models used as benchmarks for testing our method
are presented in Section  \ref{subsec:Benchmark-models}. In Section  \ref{sec:method}
we explain our proposed algorithm for using random variables for estimating
the entanglement moments of TN in general dimension, and study the
variance of the estimate in Section  \ref{sec:Error-estimation}, from
which arises the complexity of an estimation up to a chosen error.
We benchmark the algorithm with the ground states of the exactly solvable
toric code model, Eq.\ (\ref{eq:toric}), using iPEPS, and the XX chain,
Eq.\ (\ref{eq:XX-1}), using MPS, in Section  \ref{subsec:Comparison-with-theory}.
Finally, we discuss the results and future steps in the conclusions,
Section  \ref{sec:Conclusions-and-Outlook}. In the appendix, we present
variance estimations of the R\'enyi moments in the general case (Appendix
\ref{sec:appVar}), as well as specifically in the toric code model,
which is used as a benchmark (Appendix \ref{sec:appToricVar}). 

\section{\label{sec:Preliminaries}Preliminaries}

\subsection{Entanglement measures\label{subsec:Entanglement-measures}}

For a quantum 
system in a pure state 
$\rho=\k{\psi}\b{\psi}$, we define 
for a subsystem the reduced quantum state or
\emph{reduced
density matrix} (RDM) as
\begin{equation}
\rho_{A}: =\mathrm{Tr}_{\overline{A}}(\rho).\label{eq:RDM}
\end{equation}
The entanglement of the subsystem $A$ with its environment (consitutung its complement)
$\overline{A}$ is encoded in the RDM. We introduce the moments of
the RDM. For a positive integer $n$, 
the $n$-th RDM moment is defined to be
\begin{equation}
p_{n}(\rho_{A}):=\mathrm{Tr}(\rho_{A}^{n}),\label{eq:moment}
\end{equation}
also referred to as the \emph{R\'enyi moments}. On top of being entanglement monotones \citep{Wilde} and hence measures of entanglement in their own right,
these moments are used for defining various entanglement measures
with useful mathematical properties \citep{measures_1996,measures_1997,measures_1998,measures_2001,measures_2001_horodecki,measures_2002,measures_2010,measures_2021}.
The RDM moments can -- under mild mathematical conditions -- be analytically continued to the 
entanglement measure featuring the strongest interpretation for pure bi-partite quantum states, the 
\emph{von Neumann entanglement
entropy}\citep{VN} defined as
\begin{equation}
S(\rho_{A}) :=-\Tr\left(\rho_{A}\log\rho_{A}\right)
\end{equation}
for RDMs $\rho_{A}$, as the $1$-\emph{R\'enyi entropy}. The von Neumann entropy is obtained in the limit
$S(\rho_{A})=\lim_{n\rightarrow1}
(1-n)^{-1}\log\left(p_{n}(\rho_{A})\right)$.
The R\'enyi moments are especially popular as entanglement indicators since they
do not require a full reconstruction of the RDM spectrum. Therefore,
they are often easier to either calculate theoretically or measure
experimentally than other entanglement measures\citep{pastMeas_2012_entropy,Daley,
pastMeas_2015_entropy,IslamGreiner,PichlerPRX,
pastMeas_apr_2019_experimental_purity,pastMeas_feb_2018_entropy_spin,
CGSMeas_jun_2019,pastMeas_jul_2020_entropy_experiment,pastMeas_jun_2020}.

The measures above are appropriate when quantifying the entanglement
between a subsystem $A$ and its environment $\overline{A}$ when
$A\cup\overline{A}$ is in a pure state. When characterizing the entanglement
between two subsystems $A_{1}$ and $A_{2}$ whose union $A$ is not necessarily pure, the quantities 
above will no longer be suitable to quantify entanglement, as they cannot
distinguish between the quantum entanglement between $A_{1}$ and
$A_{2}$ from their entanglement with the environment. One of the best known 
measures for the entanglement between
two subsystems labeled as 
$A_{1}$ and $A_{2}$ is the
\emph{entanglement negativity} 
\cite{EisertPlenioNeg,PhD,measures_2002}, based
on the \emph{positive partial transpose
(PPT) criterion}\citep{Peres,Horodecki2_1996,Horodecki4_2009}
\begin{equation}
	\mathcal{N}(\rho_{A}):=\frac{\| \rho_{A}^{T_{2}}\|_1-1}{2},\label{eq:negativity}
\end{equation}
 where $\|\cdot\|_1$ denotes the trace norm,
and $\rho_{A}^{T_{2}}$ the partial transposition of the degrees of
freedom corresponding to $A_{2}$ in $\rho_{A}$,
\[
	\b i_{A_{1}}\b j_{A_{2}}\rho_{A}\k{k}_{A_{1}}\k{l}_{A_{2}}=
	\b i_{A_{1}}\b{l}_{A_{2}}\rho_{A}^{T_{2}}\k{k}_{A_{1}}\k j_{A_{2}},
\]
for all vectors $ (\k {i},\k{k} ), (\k {j},\k{l} )$
in an orthonormal basis of the Hilbert spaces of $A_{1},A_{2}$, respectively.
The usefulness of the negativity as an entanglement measure for two
subsystems in a mixed state 
\cite{PhD,measures_2002}
leads us to define the moments of the
partially-traced RDM, further referred to as 
\emph{PT moments}. The $n$-th
PT moment is defined to be
\begin{equation}
R_{n}(\rho_{A}) :=\mathrm{Tr}\text{\ensuremath{\left((\rho_{A}^{T_{2}})^{n}\right)}}\label{eq:renyineg}
\end{equation}
for a positive integer $n$.
The negativity can be obtained by an analytic continuation of the
PT even integer moments by $\|\rho_{A}^{T_{2}}\|_1=\lim_{n\rightarrow1/2}R_{2n}(\rho_{A})$.
The PT moments are not entanglement monotones, but they can be used to detect entanglement between $A_{1}$ and $A_{2}$\citep{negativity,negativity21},
as well as for estimating the negativity\citep{negativityML}. The
popularity of the PT moments as entanglement indicators stems from
the fact that they too do not require a full reconstruction of the
partially transposed RDM, and are therefore easier to calculate and
measure experimentally\citep{CGSMeas_jun_2019,negativity}.

For systems with a conserved charge $Q$, the
quantum state of the full system
typically commutes with the charge operator, 
\begin{equation}[\rho,\hat{Q}]=0.
\end{equation}
A partial trace can be applied to the permutation relation above to
give $[\rho_{A},\hat{Q}_{A}] = 0$, where $\hat{Q}_{A}$ is the charge
operator on subsystem $A$. The RDM is thus composed of blocks, each
corresponding to a charge value $q$ in subsystem $A$, as illustrated
in the inset of Fig.~\ref{fig:symresolved}. We denote the RDM block
corresponding to charge $q$ by $\rho_{A}(q)$. The entanglement measures,
and specifically the RDM moments, can then be decomposed into suitable contributions
from the different blocks called charge-resolved or symmetry-resolved
moments\citep{sr_Laflorencie_2014,sr_GS_PhysRevLett.120.200602,sr_PhysRevB.98.041106,sr_PhysRevLett.121.150501},
\begin{equation}
p_{n}(\rho_{A},q)
:=\Tr\left(\rho_{A}(q)^{n}\right),\label{eq:charge_resolved}
\end{equation}
again for positive integers $n$.
This definition could be extended to
the negativity as well\citep{sr_CGS_PhysRevA.98.032302}.
The study of symmetry-resolved entanglement has drawn much interest
lately, both analytically and numerically\citep{sr_PhysRevB.99.115429,sr_PhysRevB.100.235146,
sr_Fraenkel_2020,sr_10.21468/SciPostPhys.8.3.046,
sr_Capizzi_2020,
sr_PhysRevB.103.L041104,
sr_PhysRevB.101.235169,sr_PhysRevB.102.054302,
sr_PhysRevLett.125.120502,
sr_PhysRevB.102.235157,
sr_10.21468/SciPostPhys.10.3.054,
sr_2021JHEP...07..030Z,
vitale2021symmetryresolved,
negativity21,Fraenkel_symresolved_2021}
as well as in the development of experimental measurement protocols\citep{sr_GS_PhysRevLett.120.200602,
sr_CGS_PhysRevA.98.032302,doi:10.1126/science.aau0818,CGSMeas_jun_2019}.
It reveals the relation between entanglement and charge and can point
to effects such as \emph{topological phase transitions}\citep{sr_PhysRevB.99.115429,sr_Fraenkel_2020,sr_PhysRevLett.125.120502}
or to instances of \emph{dissipation} in open systems dynamics\citep{vitale2021symmetryresolved}. 

The estimation of symmetry-resolved entanglement can be done based
on the analysis in Ref.~\onlinecite{sr_GS_PhysRevLett.120.200602}:
We introduce the so-called 
\emph{flux-resolved RDM moments} as
\begin{equation}
p_{n}(\rho_{A},\varphi)
: =\Tr\left(e^{i\varphi\hat{Q}_{A}}\rho_{A}^{n}\right),\label{eq:flux_resolved}
\end{equation}
 where $\varphi\in [0,2\pi)$ can be thought of as an Aharonov-Bohm flux inserted
in the replica trick. 
The symmetry-resolved moments can be extracted
from the flux-resolved moments by a Fourier transform according to
\begin{equation}
p_{n}(\rho_{A},q)
:=\frac{1}{N_{A}}\sum_{\varphi}p_{n}(\rho_{A},\varphi)e^{-iq\varphi},\label{eq:fourier}
\end{equation}
where $\varphi=2\pi k/N_{A}$ and $k=0,\dots, N_{A}-1$. 

\begin{figure*}[!tph]
\noindent \includegraphics[width=.7\linewidth]{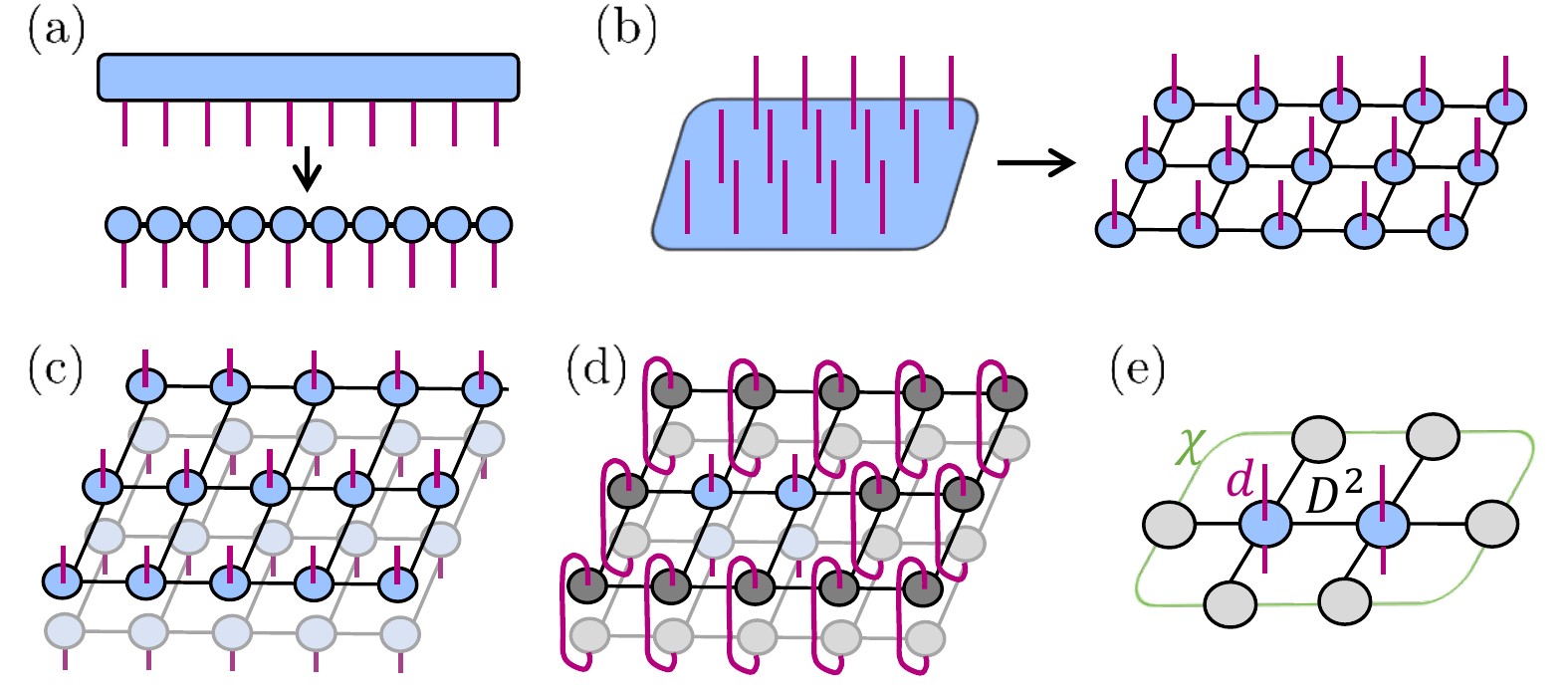}\caption{\label{fig:pepsbasics}(a-b) A general pure quantum state can be thought
of as a rank $N$ tensor represented by a single tensor with $N$
legs (indices), each with degree $d$. 
The full tensor capturing the entire quantum state can be
decomposed into $N$ tensors, each representing a single site. The
legs of the original tensor (in pink) are now divided between all
tensors, and will be referred to as the physical index. Each tensor
has additional legs connecting it to its neighbours, which will be
referred to as the bond index. Contracting all the bond indices will
result in the original tensor. (a) and (b) represent states in a one
and two spatial dimensions, respectively. (c) Representing the density
matrix of a pure state by taking two copies and applying complex conjugation
to one of them, which is indicated by the physical index facing downwards.
(d) The RDM of the sites in blue is obtained by tracing out the degrees
of freedom of the sites in grey, which in turn is obtained by contracting
the physical legs of both copies. (e) The RDM can be represented by
the site tensors of the sites in $A$ from both copies, which is here
represented by a single tensor with two physical legs, and boundary
tensors, which have only bond indices. In order to obtain the boundary
tensors, we use the boundary MPS method\citep{BMPS}. The dimension
of the bond indices in black is $D^{2}$, while that of the legs in
green, connecting the boundary tensors to each other, is $\chi$,
the bond dimension of the environment.}
\end{figure*}

\subsection{Tensor networks\label{subsec:TN}}

We will now briefly review the TN tools that 
are being used to compute the quantities presented above in the remainder of this work by means of sampling techniques.

\subsubsection{Matrix product states\label{subsec:MPS}}

We here consider a one-dimensional spin system, so of a finite local dimension,
featuring $N$ lattice sites.
The state vector of the
system can be written as 
\begin{equation}
%
\k{\psi}=\sum_{\sigma_{1},\dots,\sigma_{N}}
\Psi_{\sigma_{1},\dots,\sigma_{N}}\k{\sigma_1,\dots,\sigma_{N}},\label{eq:psi}
\end{equation}
where $\Psi$ is the rank $N$ tensor of the coefficients of $\k{\psi}$.
$\Psi$ has $d^{N}$ complex amplitudes, where $d$ is the Hilbert
space size of a single spin. In an MPS representation we decompose
$\Psi$ into $N$ different tensors, each corresponding to a single
site, as illustrated in Fig.~\ref{fig:pepsbasics}a. Each such tensor
will have a single index corresponding to the indices of the original
tensor, often called the `physical leg' or the `physical index', 
and two additional indices connecting with the tensors corresponding
to the site's neighbours, often called `entanglement legs' or `bond
indices' (with only one bond index for the sites at the edges).
Contracting all the bond indices will result in the original tensor
$\Psi$. In many cases one can limit the dimension of the bond index
to be some chosen constant $D$, also known as the bond dimension,
and discard the least significant variables. In this way, the number
of real parameters will be scaling as $O(ND^{2}d)$, at the cost of getting an approximate representation for the state. States that are expected to be 
well approximated
by such limited tensors obey an entanglement area-law\citep{TN_2008,TN_2010,TN_2013,Orus-AnnPhys-2014}
(in fact, this is provably the case for
area laws of suitable R\'enyi entropies
\cite{Schuch_MPS}).
This MPS decomposition is a widely used method for the simulation of ground states\citep{WhitePRL,
VidalMPSGS}, thermal states\citep{MPSthermal04_4,MPSthermal04_5,
MPSthermal21}
and states undergoing a
time evolution\citep{TEBD03,TEBD04,
TEBD08,
TEBD19}
generated by local Hamiltonians of  one-dimensional  systems.

For a system partitioned into two contiguous subsystems, the extraction
of the spectrum of the RDM, also called the \emph{entanglement spectrum},
is very natural\citealp{SCHOLLWOCK_2011} and can often be useful in classifying phases of matter in one spatial dimension\citep{Pollmann_ES2010,1010.3732,Pollmann_SPT2012,Kshetrimayum_SPT2015}. 
We note that a decomposition of the system into two tensors, one corresponding to subsystem $A$
and one to $\overline{A}$, is built into the decomposition of the
systems into site tensors, and that this decomposition can be transformed
into the Schmidt decomposition of the state vector
\begin{equation}
\k{\psi}=\sum_{i}\psi_{i}\k i_{A}\k i_{\overline{A}},\label{eq:schmidt}
\end{equation}
where $\{\k i_{A}\},\{\k i_{\overline{A}}\}$ are orthonormal bases
of $A,\overline{A}$, respectively. The values $\{\psi_{i}\}$ are
called the \emph{Schmidt values}, and can be extracted by a singular value decomposition\citealp{SCHOLLWOCK_2011}.
The RDM is thus
\begin{equation}
\rho_{A}=\sum_{i}|\psi_{i}|^{2}\k i_{A}\b i_{A}.\label{eq:schmidtRho}
\end{equation}
The RDM eigenvalues are thus the squared absolute values of the Schmidt
values, and by obtaining them, we can extract the RDM moment in all
ranks $n$, as well as the von Neumann entropy. Specific techniques
have been developed for the extraction of entanglement measures in
some additional cases, such as the entanglement of a contiguous subsystem
of an infinite system\citep{ciracReview} or the negativity of two
contiguous subsystems\citep{RuggieroNeg}.

\subsubsection{Projected entangled pair states\label{subsec:PEPS-1}}

For two or higher dimensional
lattice systems, the MPS formalism is extended
to an ansatz called \emph{projected entangled pair states} (PEPS)\citealp{peps_2004,peps_2008}.
The 
tensor capturing the state vector of the entire lattice
is then decomposed into $N$ site tensors, each with a single physical index and a bond index for each neighbour of the
site in the system. An example for a square lattice is depicted in
Fig.~\ref{fig:pepsbasics}b, and the generalization to other lattice
configurations is straightforward.

The \emph{infinite version of PEPS}, known as iPEPS\citep{peps_2008}, can
be used to represent states in the thermodynamic limit in 2D. They
are defined by a finite set of tensors repeated all over the lattice
with some periodicity. iPEPS have found numerous applications in studying
ground states\citep{CorbozHubbard,
Xiangkhaf,PicotPRBB2016,KshetrimayumXXZ,KshetrimayumCacro,BoosPRB2019},
thermal states\citep{CzarnikfinT2012,
KshetrimayumfinT2019,
Mondal2020}
and non-equilibrium problems\citep{KshetrimayumNatcomm2017,Czarnikevolution2019,Hubig2019,Kshetrimayum2DMBL,Kshetrimayum2DTC,
Dziarmaga2dMBL2021}
in two spatial dimensions, and have become state of the art numerical technique for studying
strongly correlated two-dimensional problems. The technique does not suffer from
the infamous sign problem\citep{EisertFPEPS,CorbozfermionPEPS} and
can go to very large system sizes, thus allowing access to regimes
where techniques like Quantum Monte Carlo and exact diagonalization
fail. 

The pure quantum state $\rho=\k{\psi}\b{\psi}$ 
of the quantum system
can be
obtained by taking a PEPS state vector and its 
Hermitian
conjugate and placing
them back to back as a tensor product, as depicted for PEPS 
in 
Fig.~\ref{fig:pepsbasics}c. We now examine a rectangular subsystem $A$
with $N_{A}=w_{1}\times w_{2}$ sites, where $w_{1}\le w_{2}$ (as
will be the notation throughout this work). In order to get the RDM of $A$ as defined in Eq.\ (\ref{eq:RDM}), the degrees of freedom
of $\overline{A}$ need to be traced out. This can be obtained by
contracting the physical legs of all tensors corresponding to sites
in $\overline{A}$ with the physical legs of the same tensor in the
complex conjugate. We get a RDM composed of site tensors for the tensors
in $A$, and boundary tensors resulting from the tensors in $\overline{A}$,
as depicted in 
Fig.~\ref{fig:pepsbasics}e. Such boundary tensors
can be approximately computed for an infinite system. We remark here
that exactly contracting PEPS tensors is a computationally hard problem (in worst case complexity and for meaningful probability measures even in average case)\citep{Schuchpepshard,Haferkampspepshard}
and therefore, we have to rely on approximation algorithms such as
the \emph{corner transfer matrix renormalization group} algorithm\citep{ctmrgNishino,ctmrgOrus2009}
\emph{boundary MPS techniques}\citep{BMPS}, 
\emph{higher order tensor renormalization
group} methods\citep{hotrg} or others. It is also known that those PEPS that are ground states of uniformly gapped parent Hamiltonians -- which are interesting in the condensed matter context -- can actually be contracted in quasi-polynomial time
\cite{samplingGappedPEPS}.
In this work, we make use
of the boundary MPS technique: We create a one-dimensional  TN representing the boundary
of the (supposedly infinite) system, and multiply it by the `traced
out' tensors indicated in 
Fig.~\ref{fig:pepsbasics}d. The boundary
bond dimension is limited to a constant dimension $\chi$. This process
is then repeated until the  one-dimensional  boundary tensors are converged, resulting
in a  one-dimensional  boundary as depicted in 
Fig.~\ref{fig:pepsbasics}e.

\subsection{Entanglement measures computed from reduced states\label{subsec:Entanglement-measures-in}}

The entanglement measures presented in Section  \ref{subsec:Entanglement-measures}
can be extracted for two contiguous systems in an MPS as presented
in Section \ref{subsec:MPS}, as well as in additional specific cases
in one\citep{RuggieroNeg,ciracReview} and two\citep{orusTopo}
spatial dimensions. However, for a general dimension and partition, there is no efficient way known to quantify the entanglement. A straightforward method
can be contracting the tensors such that the RDM is obtained explicitly to then obtain its
spectral decomposition.
However, the explicit RDM is of 
dimension $d^{N_{A}}\times d^{N_{A}}$, 
which comes along with substantial computational effort and which imposes a strong restriction on the accessible system sizes.

\begin{figure}[t]
\includegraphics[width=.7\linewidth]{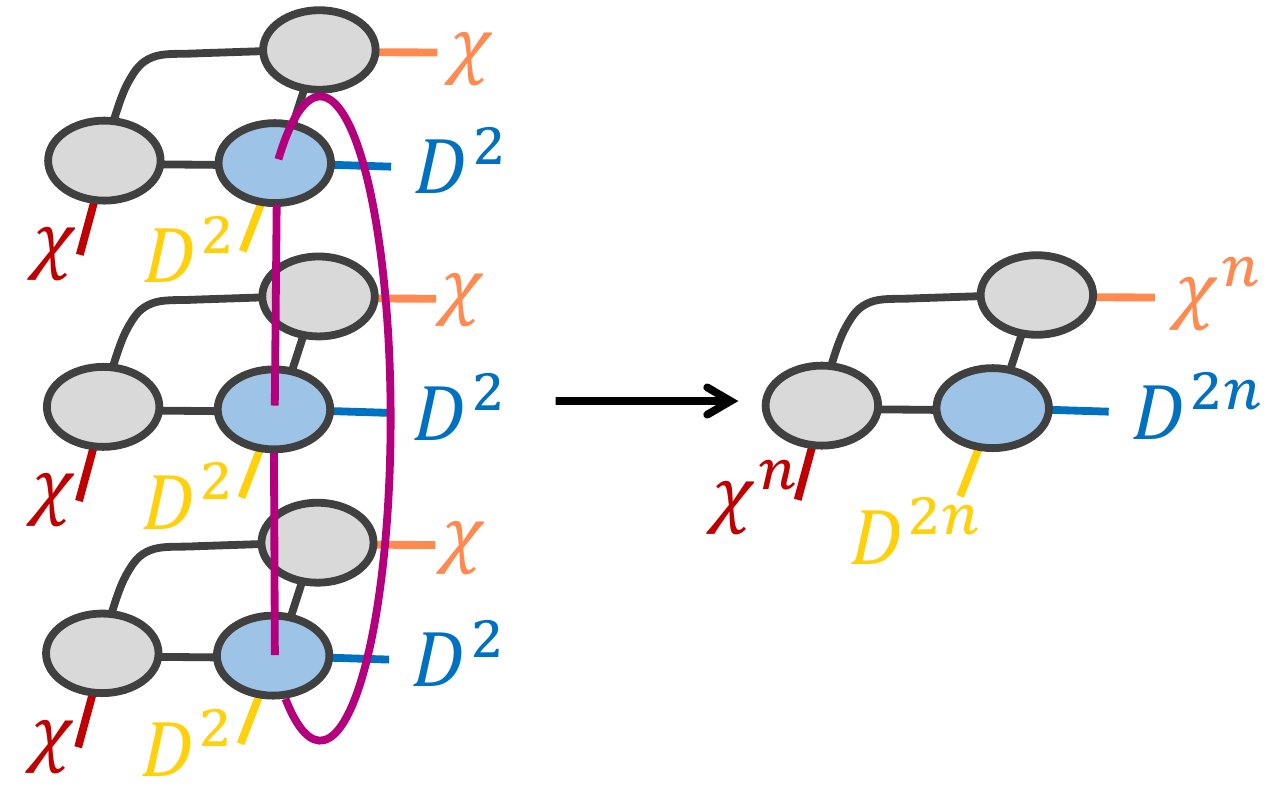}\caption{Contracting $n$ copies of the density matrix (here $n=3$) to get
the $n$-th RDM moment can be done site by site as in the figure. The
exponential dependence of the dimension of $\rho_{A}$ on $N_{A}$
turns into a linear dependence. 
However, a
prohibitive 
exponential
dependence on $n$ emerges instead.}

\label{fig:pepsreplica}
\end{figure}

That being said, the $n$-th RDM or PT moments defined in 
Eqs.~(\ref{eq:moment}, \ref{eq:renyineg})
can be calculated in polynomial time in $N_{A}$ using $n$ copies of the system tensors as depicted in 
Fig.~\ref{fig:pepsreplica}
for a two-dimensional PEPS. The space complexity required for performing this multiplication
for an MPS scales as $O\left(d^{2n}D^{2n}+D^{4n}\right)$. The space complexity
for a two-dimensional PEPS is given by $O\left(\chi^{2n}D^{2(w_{1}+1)n}+D^{8n}d\right)$,
where $\chi$ is the bond dimension of the environment as depicted
in Fig.~\ref{fig:pepsbasics}e, and $w_1$ is the short edge of the rectangular system, as define in Section \ref{subsec:PEPS-1} above. For $n>1$, the 
exponential
dependence of the cost on $n$ quickly makes it prohibitively large,
despite the fact that for a narrow system (constant $w_{1}$),
the time complexity is linear in $N_{A}$ and the space complexity
does not depend on $N_{A}$ (except for the possible dependence of
$D$ on $N_{A}$, as can sometimes happen in finite systems). 

\subsection{Benchmark models\label{subsec:Benchmark-models}}

Before turning to presenting the actual sampling method for computing entanglement measures in
quantum systems captured by tensor networks,
 we here first present the models we used for benchmarking our method: The two-dimensional gapped toric code model on a square lattice and the one-dimensional gapless XX model.

\subsubsection{The toric code model}

\begin{figure}[t]
\includegraphics[width=1\linewidth]{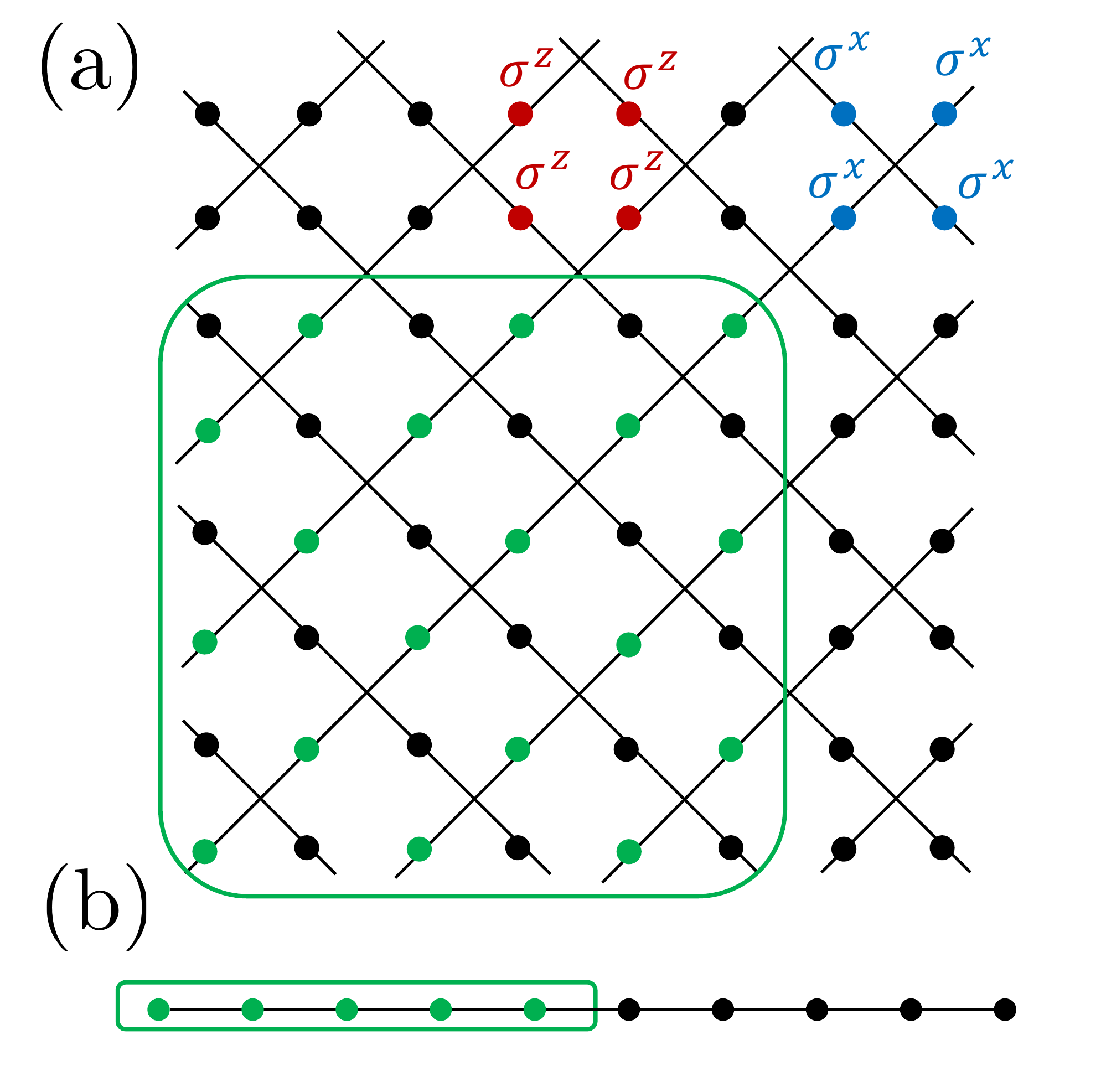}

\caption{(a) The toric code model on a square lattice. In blue is an example
of a star operator, $\otimes_{i\in s}\sigma_{i}^{x}$, and in red
an example of a plaquette operator, $\otimes_{i\in p}\sigma_{i}^{z}$.
Circled in green is an example subsystem of dimensions $w=6$, where the system is composed of the sites colored in green.
(b) The XX model is defined on a finite  one-dimensional  chain, where subsystem
$A$ is its left half as circled and colored in green.}
\label{fig:toric}
\end{figure}

The first benchmark model we elaborate on is the analytically solvable toric code model on a square lattice. The \emph{toric code}, introduced by Kitaev\citep{toric}, transferring insights from
topological quantum field theory to the realm of 
quantum spin systems, is a model of spins on a square lattice with local dimension $d=2$. The spins live on the edges of the lattice rather than its
nodes. The Hamiltonian of the model is given by
\begin{equation}
H=-J_{s}\sum_{s}\otimes_{i\in s}\sigma_{i}^{x}-J_{p}\sum_{p}\otimes_{i\in p}\sigma_{i}^{z}.\label{eq:toric}
\end{equation}
$s$ in the equation above represents the set of edges around a single
node in the lattice (a star) and $p$ represents the set of edges forming a plaquette in the lattice, as shown in Fig.~\ref{fig:toric}. The ground state of toric code model displays several important properties, among which are  topological order, which leads to robustness to
local errors, making it an important candidate for fault-tolerant
error correction code. For $J_{s},J_{p}>0$, the ground state vector
of the model with open boundary conditions
is known 
and can be written as
\begin{equation}
\k{\psi_{0}}=\prod_{s}\frac{(\mathbb{I}+\otimes_{i\in s}\sigma_{i}^{x})}{2}\k 0^{\otimes N},\label{eq:gs}
\end{equation}
where $N$ is the number of all sites in the system. 
In the limit $N\rightarrow\infty$,
the iPEPS representation of the infinite toric code ground state is
given in Refs.~\onlinecite{pepsarealaw,Orus-AnnPhys-2014}. A set of two site
tensors, $T_{A}$ and $T_{B}$, are repeated infinitely such that
all of the nearest-neighbours of a site represented by $T_{A}$ are
of the form $T_{B}$ and vice versa. The bond dimension of all bond
indices of $T_{A}$ and $T_{B}$ is $D=2$. 

For a subsystem of the infinite system in the state defined in Eq.
(\ref{eq:gs}), the density-matrix-based measures can be analytically
calculated\citep{toricentropy}. This relies on the symmetry of the
ground state under the application of $\otimes_{i\in s}\sigma_{i}^{x}$
for all stars $s$ and of $\otimes_{i\in p}\sigma_{i}^{z}$ for all
plaquettes $p$. Due to this symmetry, the RDM is block diagonal,
where the size of each block equals the order of the group generated
by each operator, which is 2 for the operators above. Considering the
fact that all non-zero blocks are identical, as can be seen from 
Eq.~(\ref{eq:gs}), the eigenvalues of the RDM can be extracted analytically.
Note that the symmetry mentioned above is not utilized in the numerical
method, so as to make our performance results applicable to general
analytically-unsolvable models, which do not posses such local symmetries.
Here, we estimate the 2nd, 3rd and 4th RDM moments, as well as the
3rd PT moment, 
for a checkerboard-like partition of a square subsystem ($w_1=w_2=w$), as shown for $w=6$ in Fig.~\ref{fig:toric}. We study the cases of $w=4, 6, 8$. Note that while the toric code displays an area law type entanglement structure, here the entire system is in the area and therefore a volume law is reached. Such
extensive partitions were shown to be interesting for the study of
topological phases in 
Refs.~\citep{partitions_2014,partitions_2015}
and following works (a more traditional geometry is studied in Appendix \ref{sec:appToricVar}).
For such systems, the $n$-th RDM moment
is shown to be\citep{toricentropy}
\begin{equation}
p_{n}(\rho_{A})=R_{n}(\rho_{A})=2^{-(w^2/2 - 1)(n-1)}.\label{eq:toricExplicitEntropy}
\end{equation}

The log of the moment deviates from an area law by an additive constant term,
reflecting the
the \emph{topological order} of the model\citep{Kitaev06,Levin06}.
Note that for the toric code ground state $p_{3}=R_{3}$ due to
the structure of the RDM discussed above. However, we compute an estimate
for $R_{3}$ based on the generally-applicable estimator defined in
Eq.\ (\ref{eq:negest}) for completeness.

\subsubsection{The XX model}

While suited for high dimensions, we note that our method is blind to the dimensionality of the system, and will apply to  one-dimensional systems
in precisely the same way
as it would for higher dimensions. Therefore, we can use  one-dimensional  models as benchmark models for testing the system. We test our model on the
ground state of the  one-dimensional  XX model captured by the local
Hamiltonian
\begin{equation}
H=J\sum_{i}\sigma_{i}^{+}\sigma_{i+1}^{-}+\mathrm{h.c.,}\label{eq:XX-1}
\end{equation}
where $i$ stands for a site in the system. The Hamiltonian can be
seen as a 
Hamiltonian of non-interacting fermions 
by virtue of the
Jordan-Wigner transformation\citep{JW} and is thus analytically
solvable\citep{Peschel_2003}. We compute the ground state of a system
of length $2l$ and extract the 2nd, 3rd and 4th RDM moments of a
contiguous half of the system. 
In contrast to the toric code, the XX
model is gapless
in the absence of a large magnetic field
and can be well 
approximated as a conformal system. For
such systems, the RDM moments of a subsystem when the total system
is in the ground state is to a good approximation predicted to be \citep{entanglement_for_physics_2009,JinKorepin},
\begin{equation}
p_{n}(\rho_{A})\sim\frac{1+n^{-1}}{6}N_{A}^{-c(n-1/n)/6},\label{eq:cftEntanglement}
\end{equation}
where $c=1$ is the conformal charge. As opposed to the toric code
ground state RDM, which is composed of $2\times2$ blocks, the XX
ground state is not as structured, and the performance of the method
is harder to predict. As such, the XX model ground state is
a good complement to the toric code ground state in the study of the
method's performance.

\section{Method\label{sec:method}}

\begin{figure}[t]
\includegraphics[width=0.5\linewidth]{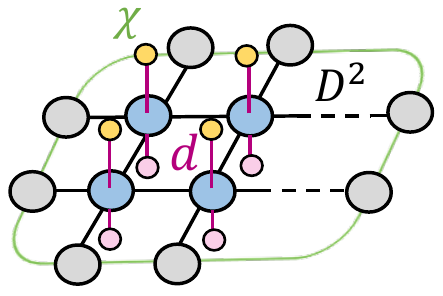}\caption{Graphical representation of the computation of a single element $\protect\b{V^{(i)}}\rho\protect\k{V^{(j)}}$
on an iPEPS. The system in the figure is a rectangle system of height
$h=2$ (and a general width $w$). The local vectors $\protect\k{v^{[\alpha]}}$
are represented by the yellow and pink one-legged circles, where the
two sets of local vectors are independent of each other.}
\label{fig:vecs}
\end{figure}

We now turn to describing the method that is at the heart of this work. The core
idea is that with suitable stochastic sampling techniques, one can more resource-efficiently estimate entanglement properties of systems captured by tensor networks.
Inspired by the growing body of 
methods based on random unitaries,
\citep{pastMeas_2010,pastMeas_2012_entropy,pastMeas_2013,pastMeas_2015_entropy,
pastMeas_feb_2018_entropy_spin,
pastMeas_mar_2019,pastMeas_apr_2019_experimental_purity,
pastMeas_jun_2019_scrambling,pastMeas_jan_2020_fidelity,
pastMeas9_may_2020_topo,pastMeas_jun_2020,pastMeas_jun_2020_scrambling,pastMeas_jul_2020_entropy_experiment,pastMeas_oct_2020_shadows}
and described in Section  \ref{sec:introduction}, we now turn to describe
our random-variables-based method for estimating the entanglement
contained in a TN state. As mentioned in the introduction, our method
differs from the protocols that are routinely implemented
in experiments by two key aspects: First,
we do not base the protocol on sampling local measurement results,
which are cumbersome to extract from TNs, but on 
sampling of expectation values,
which can be naturally calculated in TNs. Note that while 
actual sampling from
MPS can be performed exactly\citep{sampMPS}, sampling from PEPS is shown to be
computationally hard, in the worst as well as average case\citep{sampWorstPEPS,sampAveragePEPS}.
The second difference between our method and the experimental protocols
is that we do not have to limit ourselves to physically 
allowed processes, and specifically, our random operations are neither 
unitary nor quantum
channels, which allows for a significant simplification of the protocol.

\subsection{Sampling random vectors}

In what follows, random vectors $\k v \in \mathbb{C}^d$ drawn from appropriate probability measures
will feature with the property
such that
\begin{equation}
\mathbb{E} ( {\k v\b{v}} )=\mathbb{I},\label{eq:deltas}
\end{equation}
where $\mathbb{E}$ refers to the average over the
chosen probability measure.  
This is up to the normalization that is only different by a factor of $d^{1/2}$ than what is commonly called a \emph{frame} or a
\emph{spherical complex 1-design}~\cite{frames,ndesigns}. 
This convention is helpful in what follows. 
The set of vectors can be a discrete or a continuous set.
We use the random variable to get simple estimators for the entanglement
quantifiers based on R\'enyi moments of RDM of subsystems $A$, each site  $\alpha\in A$ of which
corresponds to a system of local dimension $d$. In this setting, we
consider random vectors
\begin{equation}\label{V}
\k V=\otimes_{\alpha\in A}\k {v^{[\alpha]}}\in  \mathbb{C}^{d^{N_A}},
\end{equation}
where
$\k {v^{[\alpha]}}\in \mathbb{C}^d$ are vectors drawn in an i.i.d.~fashion as  in Eq.~(\ref{eq:deltas}),
one for each
site $\alpha$. Naturally
\begin{equation}
\mathbb{E} ( {\k V\b{V}} )=\mathbb{I}\label{eq:bigdeltas}
\end{equation}
still holds true in this multi-partite setting.

\subsection{Estimators of entanglement measures}

By applying these random vectors to the reduced density
matrix, 
we obtain an estimator of the second entanglement moment,
also referred to as the purity, from expressions of the form
\begin{equation}
\hat{p}_{2}(\rho_A)=\left|\b V\rho_A\k{V^{\prime}}\right|^{2}.\label{eq:purityest}
\end{equation}
 Indeed, averaging over the (independent) random vectors $\k V,\k{V^{\prime}}$
 drawn from a product probability measure as defined in Eq.~(\ref{V}),
we consistently obtain the second moment as defined in Eq.\ (\ref{eq:moment}) as
the expectation
\begin{align}
\mathbb{E} ({\hat{p}_{2}}(\rho_A)) & =
\mathbb{E} (\b V\rho_A\k{V^{\prime}}\label{eq:purityest_explicit}
\b {V^{\prime}} \rho_A\k{V} )\\
& =\mathbb{E}
(
\Tr( \rho_A\k{V}\b V \rho_A 
\k{V^{\prime}}
\b {V^{\prime}}) )\nonumber\\
&= \Tr(\rho_A^{2})=p_2(\rho_A).\nonumber
%
\end{align}
Note again that these quantities can be readily computed at hand of the classical
description of the quantum state, but cannot be natively measured in a quantum system.
In this sense, the random sampling technique proposed here resorts to `unphysical
operations'. 

The $n$-th RDM moment can be obtained by a generalization
of Eq.\ (\ref{eq:purityest}) as the expectation 
$\mathbb{E}(\hat{p}_{n}(\rho_A))= p_n(\rho_A)$
of
\begin{equation}
\hat{p}_{n}(\rho_A)=\b{V^{(1)}}\rho_A\k{V^{(2)}}\dots\b{V^{(n)}}\rho_A\k{V^{(1)}},\label{eq:entropyest}
\end{equation}
where $\k{V^{(1)}},\dots, \k{V^{(n)}}$ are drawn in an i.i.d.~fashion from the same probability measure.
For the PT moments, 
perform the partial transposition with respect to subsystem
$A_{2}$. Specifically, we define, for $i,j=1,\dots, n$,
pairs of product vectors as
\begin{equation}
\k{V^{(i,j)}}:=\otimes_{\alpha\in A_{1}}\k{[v^{[\alpha]}]^{(i)}}\otimes_{\beta\in A_{2}}\k{[v^{[\beta]}]^{(j)}},
\end{equation}
so that the correct ordering of random product vectors can be reflected.
The estimator of the negativity moment is obtained by
\begin{eqnarray}
&&\hat {R}_{n}(\rho_A)=\b{V^{(1,n)}}\rho_A\k{V^{(2,n-1)}}\b{V^{(2,n-1)}}\rho_A\k{V^{(3,n-2)}}\nonumber
\\
&\times&
\dots
\b{V^{(n-1,2)}}\rho_A\k{V^{(n,1)}}
\b{V^{(n,1)}}\rho_A\k{V^{(1,n)}},
\label{eq:negest}
\end{eqnarray}
so that
\begin{equation}
\mathbb{E}(\hat {R}_{n}(\rho_A))= \Tr \left((\rho_A^{T_2} )^n\right)= {R}_{n}(\rho_A).
\end{equation}
Computing such an estimator on a system represented by TN is pursued
by separately computing each element $\b{V^{(i)}}\rho\k{V^{(j)}}$
or $\b{V^{(i,k)}}\rho\k{V^{(j,l)}}$, for $i,j,k,l=1,\dots, n$. The calculation
of a single element is illustrated in Fig.~\ref{fig:vecs}, and is
equivalent in terms of complexity to an expectation value calculation: For example, for a
two-dimensional PEPS, the space complexity is $O\left(\chi^{2}D^{2(w_{1}+1)}+D^{4}d\right)$,
and the time complexity is $O\left(nw_{2}\left(\chi D^{2(w_{1}+1)}\left(\chi+D^{2}\right)+D^{4}d\right)\right)$.
The calculation is repeated $M$ times over realizations of the respective
random vectors and the outcomes are averaged
in order to get an estimate 
for the desired quantity. Thus the
cost of the calculation of a single density matrix element given above,
times the number of repetitions $M$, which is discussed in Secs. \ref{sec:scaling-factors} and \ref{sec:Error-estimation}, as well as in Appendix \ref{sec:appVar}.

\subsection{Candidate probability measures}

The required property of the random vectors, captured in Eq.\ (\ref{eq:deltas}),
can be naturally obtained in a wealth of ways: After all, all that is required is to have up to 
normalization a spherical 1-design property. Still, since we do not require the
vectors to necessarily constitute a \emph{spherical complex 2-design}, the 
 second moments will depend on the specific choice of the probability measure.
For example, this can be done by
 choosing the vectors randomly out of some orthogonal
basis, or several orthogonal bases. For prime dimension $d$, 
the \emph{clock and shift operators},
\emph{Weyl operators}, or simply
\emph{$d-$dimensional
Pauli matrices}, are defined to be the 
operators
\begin{equation}
Z_{d}: =\sum_{i=0}^{d-1}\omega^{i}\k i\b i,X_{d}:=\sum_{i=0}^{d-1}\k i\b{\text{mod}(i+1,d)},\label{eq:paulis}
\end{equation}
where $\omega:=e^{2\pi i /d}.$ The $j$-th 
normalized
eigenvector of the $i$-th
$d$-dimensional Pauli matrix is denoted by $\k{p^{(i,j)}}$, and
we note that for prime $d$, the number of non-commuting Pauli matrices
is $d+1$. We then compare two possible distributions that are particularly
practical in the context given: First is the
`full-basis' distribution,
\begin{equation}
\k v\in\left\{ \sqrt{d}\k{p^{(i,j)}}\right\} _{i=1,\dots ,d+1,j=1,\dots ,d},\label{eq:distFull}
\end{equation}
The vectors are normalized
such that Eq.\ (\ref{eq:deltas}) is obeyed. The second distribution
is referred to as `partial-basis', in which we sample from the eigenbasis
of only $d$ Pauli matrices
\begin{equation}
\k v\in\left\{ \sqrt{d}\k{p^{(i,j)}}\right\} _{i=1,\dots, d,j=1,\dots ,d}.\label{eq:distPartial}
\end{equation}
For non-prime $d$, the Pauli matrices can be defined to be tensor
products of the matrices in Eq.\ (\ref{eq:paulis}) in the dimensions
of the factors of $d$. \textcolor{black}{In this case, the vector distributions defined as in Eqs.~(\ref{eq:distFull}, \ref{eq:distPartial}), but with the eigenbases of the independent products of the clock and shift operators.}
We compare the two probability measures (and
discuss why it is sufficient to only consider these distributions)
in Secs. \ref{sec:scaling-factors} and \ref{sec:Error-estimation} and in Appendix \ref{sec:appVar}
below. For states represented efficiently by TN, the partial-basis
distribution (with an optimized basis choice, as detailed in Section 
\ref{subsec:Dependence-on-basis}) turns out to be more efficient,
and therefore most of the presented results were obtained using this method.

\begin{figure}[t]
\includegraphics[width=0.9\linewidth]{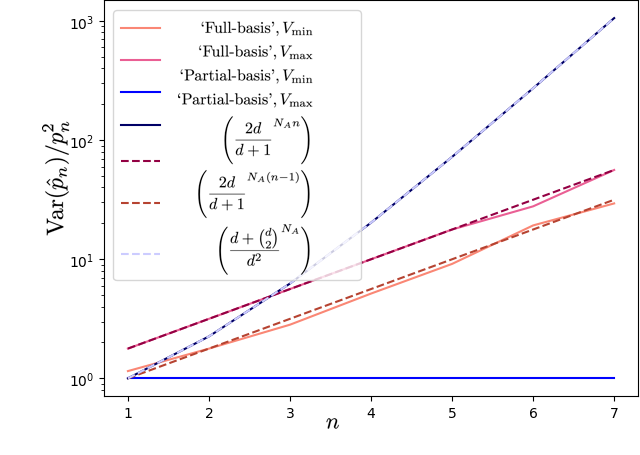}\caption{\label{fig:two_qubit}Maximal and minimal variance for a two-qubit
system using the full-basis  and partial-basis  methods, as obtained
by a gradient descent optimization. The results obtained for the maximally-
and minimally-mixed states
 in Eqs.~(\ref{eq:varianceFull1}),
(\ref{eq:varianceFull2}), and (\ref{eq:varpartial}),
are presented in dashed lines for comparison.}
\end{figure}

\subsection{Required number of repetitions}\label{sec:scaling-factors}

The method suggested makes use of random quantum states for the
estimation of entanglement measures.
When drawing random vectors from the probability measures indicated above, one finds for the probability of deviating from the expectation to be bounded by 
\begin{equation}
{\rm Pr}
\biggl(
|\hat p_n  -  \mathbb{E}(\hat p_n)  | \geq k \sigma(\hat p_n)
\biggr)\leq \frac{1}{k^2},
\end{equation}
for real $k$ and $\sigma(\hat p_n)^2:= {\rm Var}(\hat p_n)$. This is true
by virtue of \emph{Chebychev's inequality}, a large deviation bound. 
We here and in the following suppress the dependence on $\rho_A$.
For $M$
repetitions, the variance ${\text{Var}_M} $ of the estimator of the mean 
${\mathbb{E}}(\hat p_n)$  is given by
\begin{equation}
{\text{Var}_M}  :=\frac{\text{Var}(\hat{p}_n)}{M}.
\end{equation}
Since then
\begin{equation}
\varepsilon^2:= \frac{{\text{Var}_M}}{p_n^2} = \frac{\text{Var}(\hat{p}_n)}{M p_n^2},
\end{equation}
for a given $\varepsilon>0$, the number of required repetitions scales as $M = {\text{Var}(\hat{p}_n)}/( p_n^2 \varepsilon^2)$.

The characteristics of the method lead us to expect an exponential
dependence of the required number of repetitions $M$ on system size (whilst as can be seen below, a weak one). 
We thus define the scaling factor $\xi_{n}$ by 
\begin{equation}
\frac{\Var(\hat {p}_{n}(\rho_A))}{p_{n}(\rho_A)^{2}}=:\xi_{n}^{N_{A}}.\label{eq:xi}
\end{equation}
We use the notation $\xi_{n}$ for the scaling factor of $\hat R_{n}$
as well, since the scaling factors for both properties are expected
to behave similarly.

%
In Appendix \ref{sec:appVar}, we show that
the variance of the estimators defined in Eqs.~(\ref{eq:entropyest}) and (\ref{eq:negest})
is given by
\begin{equation}
\Var\left(\hat {p}_{n}\right)=\Tr((\rho_A^{\otimes2}\mathcal{E})^{n})
-p_{n}(\rho_A)^{2}\label{eq:varentropy}
\end{equation}
and
\begin{equation}
\Var(\hat {R}_{n})=\Tr\left(\left((\rho_A^{T_{2}})^{\otimes2}\mathcal{E}\right)^{n}\right)-
R_{n}(\rho_A)^{2},\label{eq:varneg}
\end{equation}
where 
\begin{equation}
{\cal E}:= \mathbb{E}\left(\k{V}\otimes \k{V}
\b{V}\otimes \b{V}\right) .
\end{equation}
Given the product structure of the probability measure, this expression is found to be
$\mathcal{E}=\otimes_{\alpha\in A}\mathcal{E}^{[\alpha]}$,
with
\begin{equation}
{\cal E}^{[\alpha]}:= \mathbb{E}\left(\k{v^{[\alpha]}}\otimes \k{v^{[\alpha]}}
\b{v^{[\alpha]}}\otimes \b{v}^{[\alpha]}\right).
\end{equation}
In a coordinate representation, this is found to be
\begin{equation}
\mathcal{E}_{i,j,kl}^{[\alpha]}=\frac{d}{d+1}(\delta_{i,k}\delta_{j,l}+\delta_{i,l}\delta_{j,k})\label{eq:generalEFull}
\end{equation}
for the full-basis distribution, and
\begin{equation}
\mathcal{E}_{i,j;k,l}^{[\alpha]}=\delta_{i,k}\delta_{j,l}+\delta_{i,l}\delta_{j,k}-\delta_{i,k}\delta_{j,l}\delta_{i,l}\label{eq:generalEPartial}
\end{equation}
for the partial-basis distribution. To give an even more specific example in a coordinate dependent form, 
for $d=2$, we have
\begin{equation}
\mathcal{E}^{[\alpha]}=\left(\begin{array}{cccc}
\frac{4}{3} & 0 & 0 & 0\\
0 & \frac{2}{3} & \frac{2}{3} & 0\\
0 & \frac{2}{3} & \frac{2}{3} & 0\\
0 & 0 & 0 & \frac{4}{3}
\end{array}\right)\label{eq:E21}
\end{equation}
and
\begin{equation}
\mathcal{E}^{[\alpha]}=\left(\begin{array}{cccc}
1 & 0 & 0 & 0\\
0 & 1 & 1 & 0\\
0 & 1 & 1 & 0\\
0 & 0 & 0 & 1
\end{array}\right)\label{eq:E22}
\end{equation}
for the full-basis and partial-basis distributions, respectively. The
variance of the symmetry-resolved moments estimator  is shown to be
bounded from above by Eq.\ (\ref{eq:varentropy}) in Appendix \ref{sec:appVar}.

\begin{figure}[!tp]
\includegraphics[width=.82\linewidth]{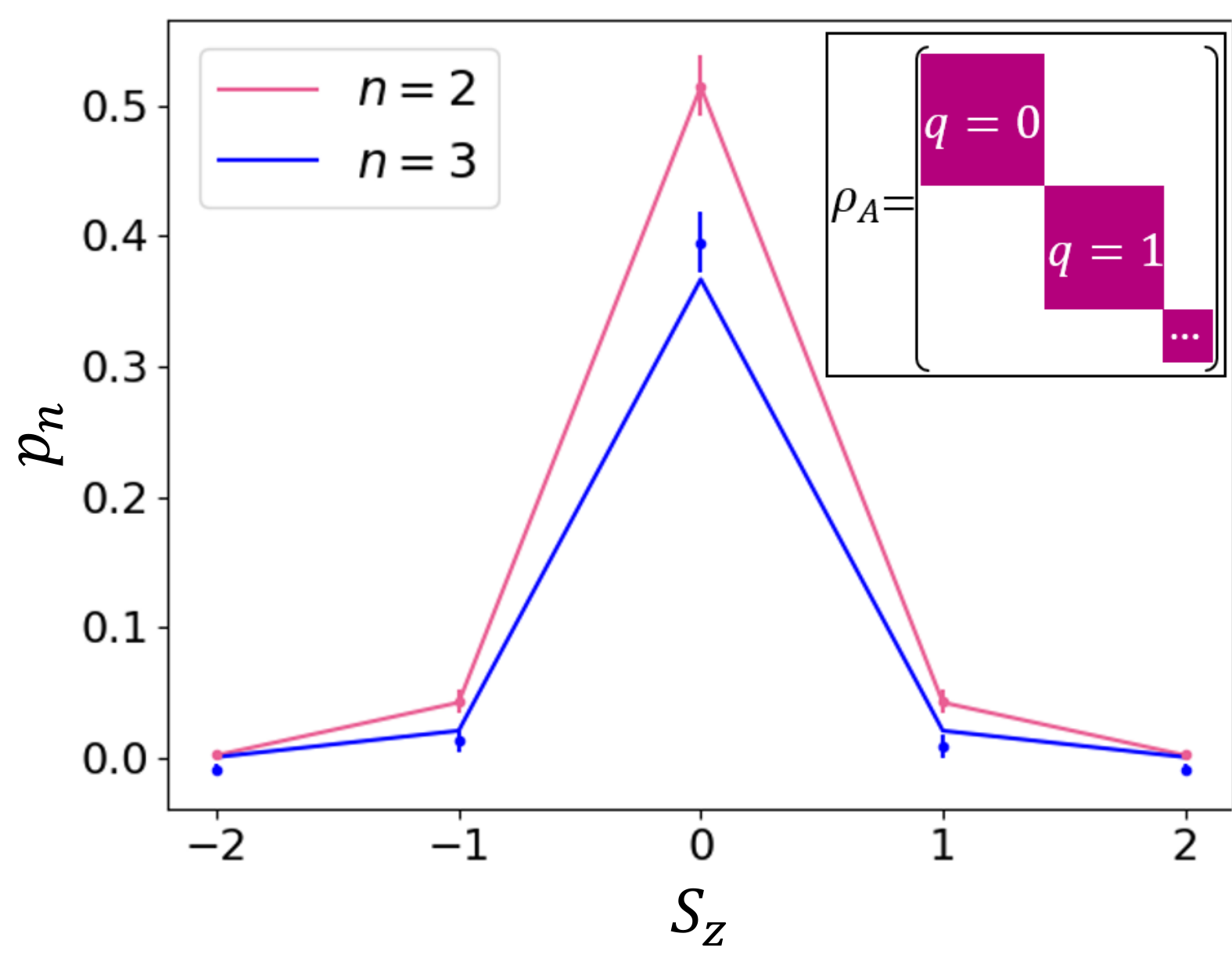}\caption{Symmetry-resolved 
RDM moments of the XX model ground state with $N_{A}=20$
and $n=2,3$. The lines are the exact values and dots are the extracted
values using the partial-basis  distribution, Eq.\ (\ref{eq:distPartial}),
with the errors estimated from the sample variance. The number of
samples $M$ used is approximately $10^{2}\xi_{3}^{N_{A}},$ where
$\xi_{3}=1.65$. Inset: When the RDM commutes with the charge operator
$\hat{Q}_{A}$, it decomposes into blocks corresponding to different
charges in subsystem $A$.}
\label{fig:symresolved}
\end{figure}

\begin{figure*}[t]
\includegraphics[width=1\linewidth]{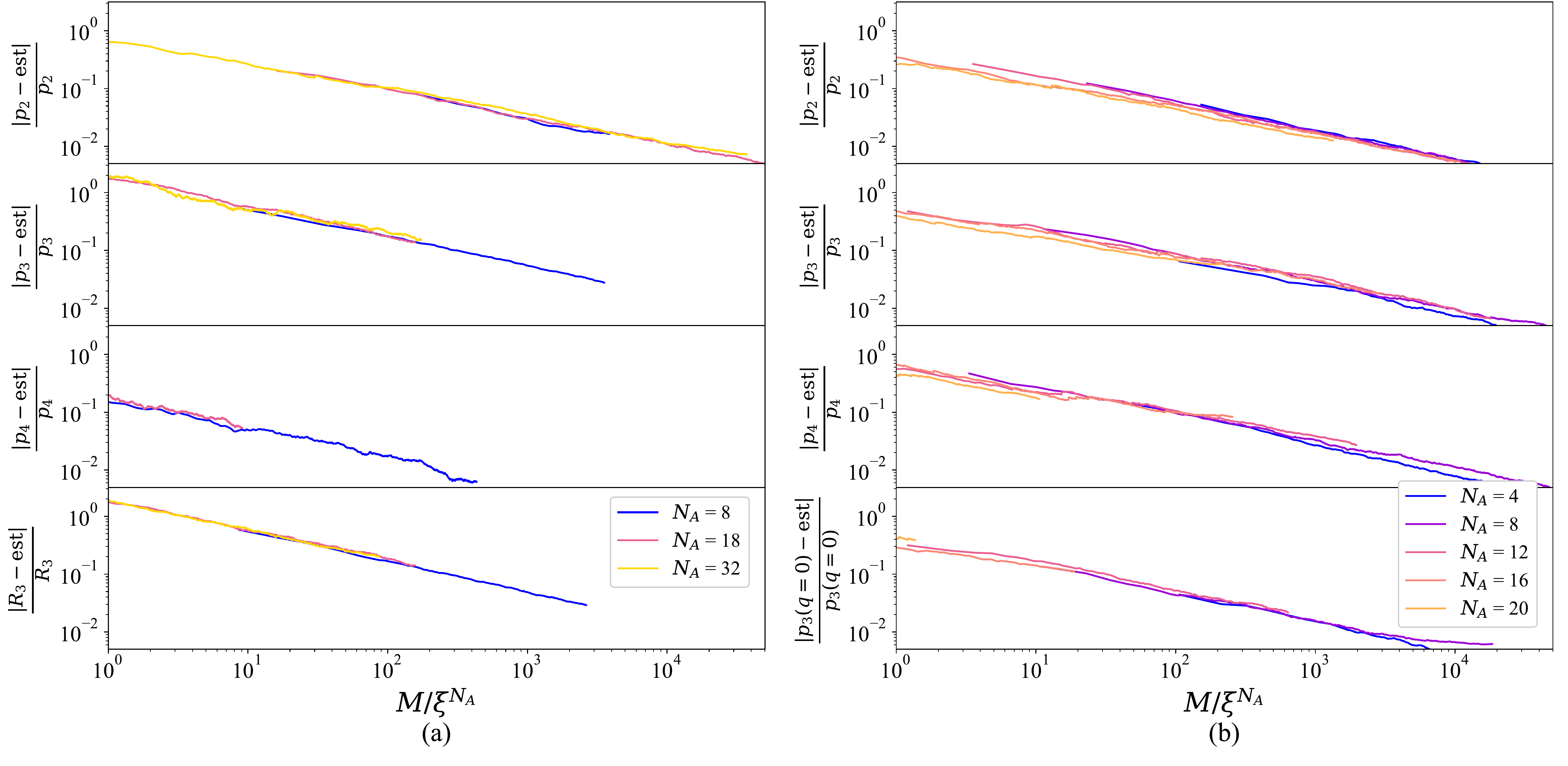}
\caption{Relative error in the estimation of (a) the 2nd-4th RDM moments and
the 3rd PT moment of the toric code ground state for a checkerboard partition ($N_A=w^2/2$), and (b) The 2nd-4th
RDM moment and the 3rd symmetry-resolved RDM moment for $q=0$ of
the XX model ground state, based on Eq.\ (\ref{eq:entropyest}) for
the RDM moments, on Eq.\ (\ref{eq:negest}) for the PT moment and (\ref{eq:fluxEst})
for the symmetry-resolved moment. Here, we use the partial-basis  distribution,
Eq.\ (\ref{eq:distPartial}). The normalization of $M$ by the scaling
factor $\xi_{n}$ in the horizontal axis is extracted from Fig.~\ref{fig:variance}
and presented in Table \ref{table:phases}. The presented plots are
averaged over 20 different permutations of the estimation results.
}
\label{fig:estimations}
\end{figure*}

We first focus on discussing the full-basis distribution. As shown in Appendix
\ref{sec:appVar}, the squared
\emph{coefficient of variation}
in a product state $\rho=\otimes_{\alpha\in A}\k{\psi}_{\alpha}\b{\psi}_{\alpha}$
and in a maximally 
mixed case (i.e., subsystem $A$ being maximally entangled with the rest of the system), $\rho_A=\mathbb{I}/d^{N_{A}}$, are
\begin{equation}
\frac{\Var(\hat {p}_{n})}{p_{n}^{2}}=\left(\frac{2d}{d+1}\right)^{nN_{A}}-1\label{eq:varianceFull1}
\end{equation}
and
\begin{equation}
\frac{\Var(\hat {p}_{n})}{p_{n}^{2}}=\left(\frac{2d}{d+1}\right)^{(n-1)N_{A}}-1,\label{eq:varianceFull2}
\end{equation}
respectively. 
The same results apply for the PT moments. For example,
for $d=2$, 
\begin{equation}
\xi_{n}=\left(\frac{4}{3}\right)^{n} 
\end{equation}
for a product state
and 
\begin{equation}
\xi_{n}=\left(\frac{4}{3}\right)^{n-1} 
\end{equation}
for a maximally 
mixed
state. Note that the scaling factors obtained below for the benchmark models,
as displayed in Table \ref{table:phases}, are in-between these two
extreme cases.

As for the partial-basis  distribution, as explained in detail in Appendix
\ref{sec:appVar}, the highest variance, hence the largest \textit{additive}
sampling error, for both the RDM moments and PT moments arises for
\begin{equation}
\rho=\otimes_{\alpha\in A}\k{\psi_{E}}\b{\psi_{E}}_{\alpha},\label{eq:pm}
\end{equation}
where $\k{\psi_{E}}$ stands for a state vector of the form 
\begin{equation}
\k{\psi_{E}}=\frac{1}{\sqrt{d}}\sum_{j=1}^{d}\left[e^{i\phi_{j}}\k j\right],
\end{equation}
with an equal magnitude of the amplitude for each state in the computational
basis. In this case, the contribution of each site to the first term
of the variance is 
\begin{equation}
\xi_{n}=\left(\left(d+4\binom{d}{2}\right)/d^{2}\right)^{n},\label{eq:varpartial}
\end{equation}
and the overall variance is given by
\begin{equation}
\left(\left(d+4\binom{d}{2}\right)/d^{2}\right)^{nN_{A}}-1.
\end{equation}
For example, for $d=2$, the scaling factor in such a case is $\xi_{n}=\left({3}/{2}\right)^{n}$.
The best case is $\rho_A=\k 0\b 0^{\otimes N_{A}}$ (or
any other basis state in the computational basis), in which the variance
is 0. Both cases are completely disentangled, and the moments equal
1. Therefore, the former of these is not maximal in terms of the
squared coefficient of variation,
which is determined by the ratio between the standard deviation
and the expected value. However, the analysis of the variance itself
already serves to demonstrate that that the choice of basis for the
vector $\k v$ can have a significant impact on the variance and due
to that on the performance of the algorithm, as discussed in Section 
\ref{subsec:Dependence-on-basis} below. The 
coefficient of variation in
the maximally nixed case in this method is calculated in Appendix
\ref{sec:appVar} and equals 
\begin{equation}
\frac{\Var(\hat p_{n})}{p_{n}^{2}}=\left(\frac{d+\binom{d}{2}2^{n}}{d^{2}}\right)^{N_{A}}.\label{eq:relativeVarPartial}
\end{equation}

The two cases represented in Eqs.~(\ref{eq:varianceFull1}),
(\ref{eq:varianceFull2}),
 and (\ref{eq:varpartial})
above, in which the variance can be calculated exactly, are not promised
to be the best or worst case for the two distribution methods. For a two-qubit system we have performed a gradient descent search for the extreme
cases in both distributions, where a basis optimization (as described
in Section  \ref{subsec:Dependence-on-basis}) has been included in the partial-basis
method. Fig.~\ref{fig:two_qubit} presents the results, which strongly
support the hypothesis that the maximally- and minimally-mixed
cases are indeed the extreme cases for the method's performance.

\begin{figure*}[!tph]
\includegraphics[width=1\linewidth]{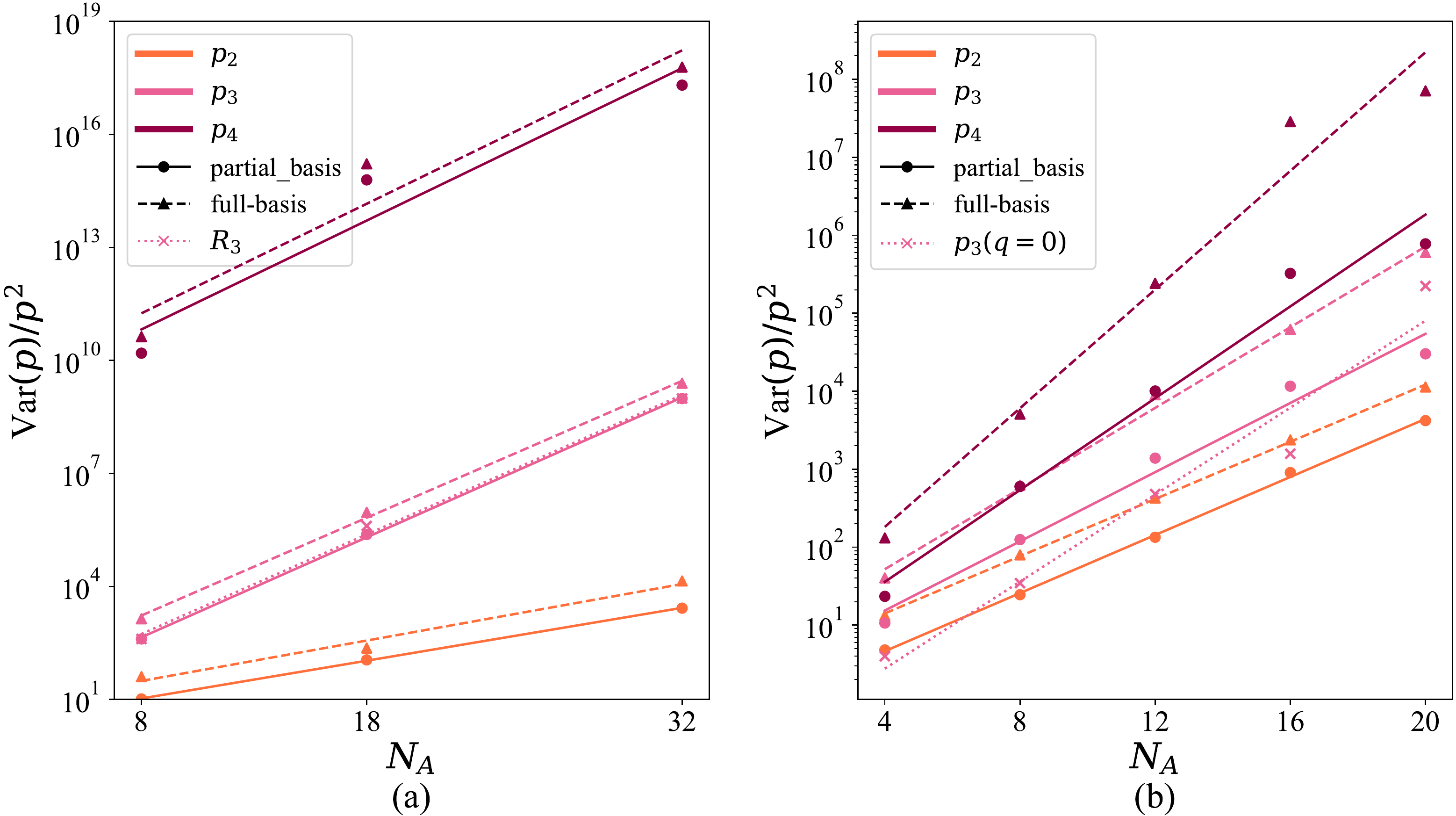}\caption{Numerical estimation of the variance of (a) the 2nd-4th RDM moments
and the 3rd PT moment of the toric code ground state for a checkerboard partition; (b) the 2nd-4th
RDM moment and the 3rd symmetry-resolved RDM moment for $q=0$ of
the XX model ground state. Here, we mainly use the partial-basis  distribution,
Eq.\ (\ref{eq:distPartial}), and add results for $p_{n}$ in the full-basis
distribution, Eq.\ (\ref{eq:distFull}), which displays a very similar
performance in the more strongly entangled toric code model, and a
worst behaviour in the log-like entangled XX model. In the vertical
axis title, $p$ stands for the estimated moment in all cases. The
dependence of the variance on system size $\xi$, defined in Eq.\ (\ref{eq:xi}),
is extracted from the linear fit in the figure and shown in Table
\ref{table:phases}. Note that for the toric code ground state shown
in (a), the behaviours of the 3rd RDM moment and the 3rd PT moment
are expected to be identical, and are slightly different only due
to the random nature of our protocol.}
\label{fig:variance}
\end{figure*}

The best distribution choice is therefore case-dependent: For moderate
or large $n$ and highly mixed cases, the full-basis  distribution
is advantageous. For small $n$ or weakly entangled cases, the partial-basis
method with a smart basis choice (as discuss in Section  \ref{subsec:Dependence-on-basis})
is more beneficial. Based on Eq.\ (\ref{eq:relativeVarPartial}), the
partial-basis seems to have a poor performance on highly entangled (mixed)
states. However, states represented efficiently by TN feature an 
entanglement that exhibits an area law, which restrains the entanglement of relevant states
to begin with. Note that even in this worst case our method is still favourable compared to the 
 the time required for an exact diagonalization of the
RDM, which scales as $O(d^{3N_{A}})$, for 
intermediate
$n$. The variances
calculated above compare favorably with the variances in the experimental
sampling-based protocols\citep{pastMeas_2010,pastMeas_2013,pastMeas_2012_entropy,pastMeas_2015_entropy,
pastMeas_feb_2018_entropy_spin,
pastMeas_mar_2019,pastMeas_apr_2019_experimental_purity,
pastMeas_jun_2019_scrambling,pastMeas_jan_2020_fidelity,
pastMeas9_may_2020_topo,pastMeas_jun_2020,pastMeas_jun_2020_scrambling,pastMeas_jul_2020_entropy_experiment,pastMeas_oct_2020_shadows,negativity},
as calculated in Ref. ~\onlinecite{negativity}: For $d=2$, the relative
variances for $n=2,3$ are shown to be $\Var(\hat{p}_{2})/p_{2}^{2}\ge(8/p_{2}^{2})\max\left\{ 2^{N_{A}}p_{2},2^{1.5N_{A}}\right\} $
and $\Var(\hat{p}_{3})/p_{3}^{2}\ge\left(39/p_{3}^{2}\right)\max\left\{ 2^{N_{A}}p_{2}^{2},2^{1.5N_{A}}p_{2},2^{2N_{A}}\right\} $.

It would be interesting to use the analysis above as a basis for a
study regarding the number of local samples required for the estimation
of R\'enyi moments in general. Naively, the moments are defined as
a function of the full RDM, which has $d^{2N_{A}}$ elements, hence
should require a comparable number of samples. However, an extraction
of a single degree of freedom is expected to require a smaller number
of samples, as is the case in small $n$s in our method. Such analysis
may point to the amount of information contained in R\'enyi moments
of different ranks.

\subsection{Symmetry-resolved RDM moments}

Using the locality of the phase operator from Eq.\ (\ref{eq:flux_resolved}),
$e^{i\varphi\hat{Q}_{A}}=\otimes_{\alpha\in A}e^{i\varphi\hat{Q}_{\alpha}}$,
we can extract the flux-resolved moments from the TN, and substitute
them into Eq.\ (\ref{eq:fourier}) to get the symmetry-resolved moments.
The estimator of the flux-resolved moment is similar to the estimator
of the full moments in Eq.\ (\ref{eq:entropyest}) 
and is obtained from
\begin{align}
\hat{p}_{n}(\rho_{A},\varphi)= & \b{V^{(1)}}e^{i\varphi\hat{Q}_{A}}\rho_{A}\k{V^{(2)}}
 \b{V^{(2)}}\rho_{A}\k{V^{(3)}}
\nonumber \\
 & \times \dots\b{V^{(n)}}\rho_{A}\k{V^{(1)}}.\label{eq:fluxEst}
\end{align}
The estimator of the charge-resolved moment is
obtained from
\begin{align}
\hat{p}_{n}(\rho_{A},q)= & \sum_{\phi}e^{-iq\varphi}\b{V^{(1)}}e^{i\varphi\hat{Q}_{A}}\rho_{A}\k{V^{(2)}}\dots\nonumber \\
 &\times  \dots\b{V^{(n)}}  \rho_{A}\k{V^{(1)}}.\label{eq:chargeEst}
\end{align}
A similar analysis of symmetry-resolved PT moments has been done in Ref.~\onlinecite{sr_CGS_PhysRevA.98.032302},
and the extension to their estimation is natural. Below we estimate
the symmetry-resolved RDM moments for the XX model and its conserved
total $S_{z}$. For this model, the symmetry-resolved moments can
be obtained exactly following Ref.~\onlinecite{sr_GS_PhysRevLett.120.200602}.
The expected and extracted results for $q\mapsto p_{n}(\rho_{A},q)$ for $n=2,3$
are displayed in Fig.~\ref{fig:symresolved}.

\section{Testing the method against the benchmark models\label{subsec:Comparison-with-theory}}

\subsection{Specific tests}
We have tested the model against the exactly solvable two-dimensional 
toric code model and  one-dimensional  XX model as detailed in Section  \ref{subsec:Benchmark-models},
employing the TensorNetwork library\citep{TN}. The precision of the
estimation for both models as a function of $M$, the number of samples
of the expressions in Eqs.~(\ref{eq:purityest}), (\ref{eq:entropyest}), (\ref{eq:negest}), and (\ref{eq:fluxEst})
is shown in Fig.~\ref{fig:estimations}. The results are obtained
using the partial-basis  distribution, Eq.\ (\ref{eq:distPartial}),
and are optimized based on the analysis in Section  \ref{subsec:Dependence-on-basis}
below. In order to reduce the numerical noise in the dependence of
the precision in $M$, we averaged this dependence over several permutations
of the $M$ repetitions. While the required number of repetitions
$M$ (for a given allowed error $\varepsilon>0$) is exponential in
system size (as discussed above in Section 
\ref{sec:scaling-factors}), it has a relatively small base $\xi_{n}$.
When considering the significant decrease in required memory space,
our method can become advantageous for systems around $N_{A}=20$,
for which the method described in Section \ref{subsec:TN} can become
too heavy in memory demands for a standard computer workstation.

\subsection{Variance estimation\label{sec:Error-estimation}}

Here we follow the analysis of the scaling factor $\xi_n$ in Section \ref{sec:scaling-factors} and estimate the scaling factors in the benchmark models, in order to get some idea regarding the variance in the general case.
The scaling factors $\xi_{n}$ obtained for
the toric code and XX ground states, for $n=2,3,4$ are estimated numerically. 
In Appendix \ref{sec:appToricVar}, we demonstrate an exact calculation of the expressions in Eqs.~(\ref{eq:varentropy}) and (\ref{eq:varneg}) for a narrow strip-like system in the toric code model, and show that the resulting expressions agree with the numerically estimated scaling factors. 

We emphasize that
the models used are not specifically suitable for the method, and
are not expected to have a low scaling factors based on Eqs.~(\ref{eq:varentropy}) and (\ref{eq:varneg}).
The estimated scaling factor $\xi_{n}$ can therefore be considered
typical.

Fig.~\ref{fig:variance} presents the estimated variances of the benchmark
models in the full-basis  distribution and partial-basis  distribution
after the basis-choice optimization detailed in Section  \ref{subsec:Dependence-on-basis}.
The scaling factor can be extracted from the dependence of the variance
on $N_{A}$. We see that the scaling factors are smaller than the
worst case presented above. In the XX model, for which the entanglement
is log-dependent in system size, we get a better performance with
the basis-optimized partial-basis  distribution. In the more strongly
entangled (and therefore less basis-sensitive) toric code model, the
difference in performance between the two methods is clearly less
significant.

\subsection{Dependence on basis choice\label{subsec:Dependence-on-basis}}

\begin{figure}[t!]
\includegraphics[width=1\linewidth]{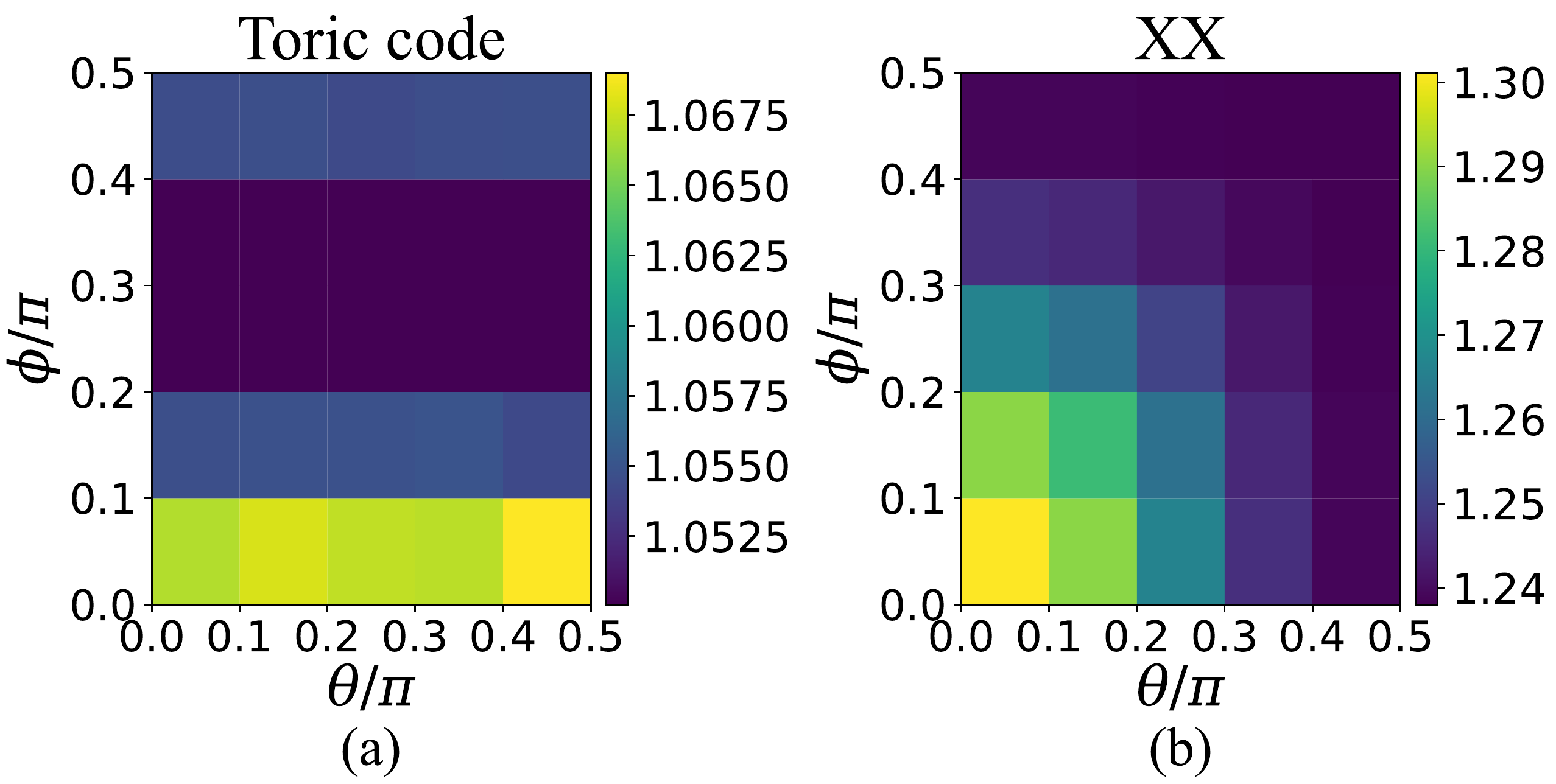}\caption{$\xi_{1}$, the scaling factor of the normalized variance of the first
moment with system size as defined in Eq.\ (\ref{eq:xi}), as a function of the working basis defined by $\phi,\theta\in [0,2\pi)$ in Eq.\ (\ref{eq:basischoice})
for the partial-basis  distribution, Eq.\ (\ref{eq:distPartial}), and
for (a) the toric code ground state for a checkerboard partition and (b) the XX model ground state.}
\label{fig:phases}
\end{figure}

In the partial-basis  distribution, as shown above, the largest and
smallest additive variance values both correspond to a completely
disentangled case, and the difference between the two stems from the
single-particle basis choice alone. This can be understood by the
decomposition 
\begin{equation}
\mathcal{E}^{[\alpha]}=\mathbb{I}\otimes\mathbb{I}+\frac{1}{2}\left(\sigma^{x}\otimes\sigma^{x}+\sigma^{y}\otimes\sigma^{y}\right),
\end{equation}
as can be seen from Eqs.\ (\ref{eq:E21}) and (\ref{eq:E22}), that demonstrates the orientation
dependence of $\mathcal{E}$ in this case (in contrast with
\begin{equation}
\mathcal{E}^{[\alpha]}=\mathbb{I}\otimes\mathbb{I}+\frac{1}{3}\left(\sigma^{x}\otimes\sigma^{x}+\sigma^{y}\otimes\sigma^{y}+\sigma^{z}\otimes\sigma^{z}\right)
\end{equation}
in the full-basis  case). For a translationally invariant system, one
can expect that the optimal basis choice for each site will be the
same. We can now attempt to decrease the variance by finding the basis
for which $\Var(\hat p_{1})$ is minimal, and use this basis for the estimation
of higher moments. We have tested this idea against the first moment
of the two benchmark models, by rotating the random vectors 
\begin{equation}
\k {v^{[\alpha]}}\longmapsto e^{i\phi\sigma^{x}}e^{i\theta\sigma^{y}}\k {v^{[\alpha]}},\label{eq:basischoice}
\end{equation}
and finding the best basis, as demonstrated in Fig.~\ref{fig:phases}.
We plot the scaling factor for $n=1$, $\xi_{1}$, for the two benchmark
models as a function of the basis choice. We then compare the best
and worst choices of $\phi,\theta\in [0,2\pi)$ and extract the variance of the
higher moments in the corresponding bases, as summarized in Table
\ref{table:phases}. We can see that the $p_{1}$ case acts as a good
indicator for the basis choice of the moments in higher $n$s, and
allows for a smart basis choice which decreases the variance.

\begin{table}[t]
\begin{tabular}{|>{\centering}p{0.19\columnwidth}|>{\centering}p{0.18\columnwidth}|>{\centering}p{0.18\columnwidth}|>{\centering}p{0.2\columnwidth}|>{\centering}p{0.17\columnwidth}|}
\hline 
 & Toric code 

best case & Toric code 

worst case & XX 

best case & XX 

worst case\tabularnewline
\hline 
$(\phi,\theta)$ & $(0, 0.3\pi)$ & $(0.5\pi,0)$ & $(0.5\pi,0.5\pi)$ & $(0,0)$\tabularnewline
\hline 
\hline 
$\xi_{1}$ & {$1.05$} & {$1.07$} & $1.24$ & $1.30$\tabularnewline
\hline 
$\xi_{2}$ & {$1.23$} & {$1.36$} & $1.54$ & $1.73$\tabularnewline
\hline 
$\xi_{2}$ 

{[}Full-basis{]} & \multicolumn{2}{c|}{{$1.26$}} & \multicolumn{2}{c|}{$1.51$}\tabularnewline
\hline 
$\xi_{3}$ & {$1.80$} & {$1.89$} & $1.65$ & $1.98$\tabularnewline
\hline 
$\xi_{3}$ 

{[}Full-basis{]} & \multicolumn{2}{c|}{{$1.81$}} & \multicolumn{2}{c|}{$1.77$}\tabularnewline
\hline 
$\xi_{4}$ & {$2.75$} & {$2.91$} & $2.09$ & $2.50$\tabularnewline
\hline 
$\xi_{4}$ 

{[}Full-basis{]} & \multicolumn{2}{c|}{{$2.81$}} & \multicolumn{2}{c|}{$2.34$}\tabularnewline
\hline 
\end{tabular}\caption{The extracted scaling factor defined in Eq.\ (\ref{eq:xi}), $\xi_{n}$,
for $n=2,3,4$ for the best and worst case in the toric code ground
state extracted from Fig.~\ref{fig:phases} and the XX model ground
state. The scaling factors obtained in the full-basis  method are displayed
for comparison. The worst and best case in $\xi_{1}$ are shown to
predict well the dependence of $\xi_{n}$ on the basis choice.}
\label{table:phases}
\end{table}

\section{Conclusions and Outlook\label{sec:Conclusions-and-Outlook}}

In high-dimensional TN states, the naive computation of entanglement
is highly sensitive to the size of the system (when explicitly extracting
the RDM) or the bond dimension of the site tensors and boundaries
(when performing the replica trick). We developed a method for estimating
RDM moments in Eq.\ (\ref{eq:moment}) and PT moments in Eq.\ (\ref{eq:renyineg})
of such systems, as well as their symmetry-resolved components in
Eqs. (\ref{eq:charge_resolved}) and (\ref{eq:flux_resolved}), without
fully reconstructing the density matrix or contracting several copies
of the state. The method uses randomization in order to correlate
separate copies of the TN state, allowing for the estimation of properties
that are defined using more than one copy of the RDM. Though we are
inspired by recent experimental protocols\citep{pastMeas_2015_entropy,pastMeas_2010,pastMeas_2012_entropy,pastMeas_2013,
pastMeas_feb_2018_entropy_spin,
pastMeas_mar_2019,pastMeas_apr_2019_experimental_purity,
pastMeas_jun_2019_scrambling,pastMeas_jan_2020_fidelity,
pastMeas9_may_2020_topo,pastMeas_jun_2020,pastMeas_jun_2020_scrambling,pastMeas_jul_2020_entropy_experiment,pastMeas_oct_2020_shadows},
we developed a completely new algorithm which is suitable to classical
simulations, takes advantage of their strengths such as ability to
estimate the expectation value of non-Hermitian operators, and avoids
their weakness in sampling the outcomes of random measurements.

We have demonstrated our method with the iPEPS representation of the
toric code ground state and the MPS representation of the XX ground
state, and compared the results with analytical calculations. The method
can be readily used for any tensor network ansatz representing a spin
or bosonic system, and provide information on the entanglement of
systems that were formerly unreachable by today's computers due to
a strong exponential dependence of the memory space in the moment
degree $n$.  Additionally, our method
is advantageous for nontrivial partitions in one or higher spatial
dimensions, such as the checkerboard partition~\citep{partitions_2014} or random partition~\citep{partitions_2015},
for which the moments are hard to calculate even for one-dimensional MPS.

We compare two options for the random distribution, where each of
the methods turns out to be suitable for different cases. For small
$n$s, the scaling of required samples number $M$ with system size
$N_{A}$ turns out to be lower than the scaling of RDM size, and can
have implications regarding the information contained in these moments.
It would be interesting to try and develop a sampling-based method
which incorporates non-physical operations and compare its performance
with ours. The analysis of such a protocol may shed more light on
the power of R\'enyi moments.

The method should be suitable to fermionic PEPS\citep{EisertFPEPS,CorbozfermionPEPS},
and can be generalized to additional R\'enyi measures, such as participation
entropies, used for the detection of many-body localization\citep{PEs}.
Exploring the possibility of derandomizing the algorithm, similarly
to the recent results of Huang \textit{et al.}\citep{derandomization}
would also be interesting. In contrast to setting of shadow estimation, 
the very quantum state is already classically efficiently represented, and 
computing overlaps with suitable random vectors gives rise to an effective
estimation of entanglement properties. 
Now that the option to use non-physical sampling has been opened, it can be expanded to various platforms, including experimental setting with a vectorized density matrix.
It is the hope that this work 
contributes to the program of exploiting the power 
of random measurements in quantum physics, even in situations where the sampling scheme itself is not reflected by physical operations.

\begin{acknowledgments}
We thank I.~Arad, G.~Cohen, and E.~Zohar for very useful discussions.
Our work has been supported by the Israel Science Foundation (ISF)
and the Directorate for Defense Research and Development (DDR\&D)
grant No.~3427/21 and by the US-Israel Binational Science Foundation
(BSF) Grants No.~2016224 and 2020072. N.~F.~is supported by the Azrieli
Foundation Fellows program. J.~E.~has been supported by the DFG (CRC 183, project B01).
This  work  has  also  received  funding  from  the  European  Union's Horizon 2020 research and innovation programme under grant agreement  No.~817482 (PASQuanS).

\end{acknowledgments}

\newpage
\appendix

\section{Full derivation of the variance \label{sec:appVar}}

Below we derive Eqs.~(\ref{eq:varentropy}), (\ref{eq:E21}) and (\ref{eq:E22})
of the main text, and use them to find density matrices which extremize
the variance. First, we write the expression for the RDM moments estimator
explicitly. In the above coordinate independent fashion, this derivation is 
straightforward. 
The 
$n$-th RDM moment is obtained as the expectation of
\begin{equation}
\hat {p}_{n}(\rho_A)=\b{V^{(1)}}\rho_A\k{V^{(2)}}\dots\b{V^{(n)}}\rho_A\k{V^{(1)}},
\end{equation}
where again, the product state vectors
$\k{V^{(1)}},\dots, \k{V^{(n)}}$ are drawn in an i.i.d.~fashion from the same probability measure.
In expectation, we find
\begin{eqnarray}
\mathbb{E}(\hat{p}_{n}(\rho_A))&=&
\mathbb{E}(
\b{V^{(1)}}\rho_A\k{V^{(2)}}\dots\b{V^{(n)}}\rho_A\k{V^{(1)}})\\
&=&\mathbb{E}(\Tr(
\k{V^{(1)}}\b{V^{(1)}}\rho_A\k{V^{(2)}}\dots\b{V^{(n)}}\rho_A))\nonumber \\
&=&\Tr(\rho_A^n) = p_n(\rho_A).\nonumber
\end{eqnarray}
Similarly, for the PT moments, 
one can make use of random vectors of the form
\begin{equation}
\k{V^{(i,j)}}=\otimes_{\alpha\in A_{1}}\k{[v^{[\alpha]}]^{(i)}}\otimes_{\beta\in A_{2}}\k{[v^{[\beta]}]^{(j)}},
\end{equation}
for $i,j=1,\dots, n$,
so that the estimator of the negativity moment is
\begin{widetext} 

\begin{eqnarray}
\hat{R}_{n}(\rho_A)&=&\b{V^{(1,n)}}\rho_A\k{V^{(2,n-1)}}\b{V^{(2,n-1)}}\rho_A\k{V^{(3,n-2)}}\label{eq:negest_appendix}
\times
\dots
\b{V^{(n-1,2)}}\rho_A\k{V^{(n,1)}}
\b{V^{(n,1)}}\rho_A\k{V^{(1,n)}},\nonumber
\end{eqnarray}
since one simply finds by performing partial transposes in all terms
\begin{eqnarray}
\mathbb{E}(\hat{R}_{n}(\rho_A))&=&\mathbb{E}\Tr(
\b{V^{(1,n)}}\rho_A\k{V^{(2,n-1)}}\b{V^{(2,n-1)}}\rho_A\k{V^{(3,n-2)}}
\\
&\times&
\dots
\b{V^{(n-1,2)}}\rho_A\k{V^{(n,1)}}
\b{V^{(n,1)}}\rho_A\k{V^{(1,n)}})\nonumber\\
&=& \Tr\left(
(\rho_A^{T_2})^n
\right),\nonumber
\end{eqnarray}
so that indeed the correct moment of the partially transposed operator is recovered.
The variance can then be calculated from the expectation of
\begin{eqnarray}
\hat p_n(\rho_A)^2 &=&
(\b{V^{(1)}}\rho_A\k{V^{(2)}})^2
(\b{V^{(2)}}\rho_A\k{V^{(3)}})^2
\dots
(\b{V^{(n)}}\rho_A\k{V^{(1)}})^2
\end{eqnarray}
from which the square of $p_n(\rho_A)$ is subtracted. The subtle point is now that projections appear twice
rather than once. This can be reflected by making use of two tensor factors. Upon reordering
the tensor entries, one immediately finds the expression
\begin{eqnarray}
\hat p_n(\rho_A)^2 &=&
\Tr\left(
( \b{V^{(1)}}\otimes \b{V^{(1)}}) 
(\rho_A\otimes \rho_A)
(\k{V^{(2)}}\otimes \k{V^{(2)}})
\dots
( \b{V^{(n)}}\otimes \b{V^{(n)}}) 
(\rho_A\otimes \rho_A)
(\k{V^{(1)}}\otimes \k{V^{(1)}})
\right)\\
&=&
\Tr\left(
\prod_{j=1}^n
(\k{V^{(j)}}\otimes \k{V^{(j)}})
( \b{V^{(j)}}\otimes \b{V^{(j)}}) 
(\rho_A\otimes \rho_A)
\right) .
\end{eqnarray}
\end{widetext}
In expectation, this is
$\mathbb{E}(\hat p_n(\rho_A)^2)=\Tr((\rho_A^{\otimes2}\mathcal{E})^{n})$
with
\begin{equation}
{\cal E}= \mathbb{E}\left(\k{V}\otimes \k{V}
 \b{V}\otimes \b{V}\right) 
\end{equation}
and hence
$\mathcal{E}=\otimes_{\alpha\in A}\mathcal{E}^{[\alpha]}$,
where
\begin{equation}
{\cal E}^{[\alpha]}= \mathbb{E}\left(\k{v^{[\alpha]}}\otimes \k{v^{[\alpha]}}\b{v^{[\alpha]}}\otimes \b{{v}^{[\alpha]}}\right) .
\end{equation}
In this way, one finds the expression for the variance
\begin{equation}
\Var(\hat {p}_{n}(\rho_A)) = \Tr((\rho_A^{\otimes2}\mathcal{E})^{n}) - p_n^2.
\end{equation}
Here it is relevant that the frames made use of do \emph{not} necessarily have to 
constitute \emph{complex spherical 2-designs} for estimating the above entanglement measures
in an unbiased fashion,
so that the average does not necessarily resemble that of the Haar average, and may depend on the ensemble.
We now see why it is meaningful to consider the two probability measures
specified above:  Drawing vectors from the eigenbasis of a single Pauli
matrix, i.e., randomly sampling RDM elements in this basis, would
give, for example if we drew from the eigenbasis of $Z$, 
\begin{equation}
\mathcal{E}_{i,k;j,l}^{[\alpha]}=\delta_{i,j}\delta_{k,l}+\delta_{i,k}\delta_{j,l}+\delta_{i,l}\delta_{j,k}-2\delta_{i,k}\delta_{j,l}\delta_{i,l}.
\end{equation}
This matrix equals the matrix obtained 
above,
with added positive terms, hence, can only increase the variance.
The same argument can be made for any number of Pauli matrices between
2 and $d-1$. The Pauli matrices constitute a \emph{unitary 1-design}
\cite{ndesigns}, which is the
requirement for them to be a universal measure for the estimators
in Eqs.~(\ref{eq:purityest}), (\ref{eq:entropyest}), (\ref{eq:negest}), and (\ref{eq:chargeEst}).
Therefore, it is unnecessary to consider additional distributions. 
In the maximally mixed 
case, $\rho_A=(\mathbb{I}_{d}/d)^{\otimes N_{A}}$,
the normalized first term is
\begin{align}
	\frac{\Tr\left(\left[\rho_A^{\otimes2}\mathcal{E}\right]^{n}\right)}{p_{n}^{2}} & =\text{\ensuremath{\left(\frac{2d}{d+1}\right)}}	^{(n-1)N_{A}} 
\end{align}
for the full-basis and 
\begin{align}
	\frac{\Tr\left(\left[\rho_A^{\otimes2}\mathcal{E}\right]^{n}\right)}{p_{n}^{2}} & =
	\left(\frac{d+\binom{d}{2}2^{n}}{d^{2}}\right)^{N_{A}}
\end{align}
for the partial-basis, respectively. In
a product state, $\rho_A=\otimes_{\alpha\in A}\k{\psi}_{\alpha}\b{\psi}_{\alpha}$
in the full-basis  distribution, one finds
\begin{align}
\frac{\Tr\left(\left[\rho_A^{\otimes2}\mathcal{E}\right]^{n}\right)}{p_{n}^{2}}
=\text{\ensuremath{\left(\frac{2d}{d+1}\right)}}^{nN_{A}}. 
\end{align}
The performance of the partial-basis  distribution for a product state
can be analyzed as follows: $\mathcal{E}^{[\alpha]}$ is a block diagonal
matrix, with $d$ blocks of the form $B_1=(1)$ and $\binom{d}{2}$
blocks of the form 
\begin{equation}
B_2= \left(\begin{array}{cc}
1 & 1\\
1 & 1
\end{array}\right). 
\end{equation}
The largest eigenvalue of $\prod_{\alpha\in A}\mathcal{E}^{[\alpha]}$
is thus $2^{N_{A}}$, and corresponds to an eigenvector of the form
$\otimes_{\alpha\in A}\k{i,j}$, where $i$ and $j$ correspond to
the spin of the site $\alpha$ in the two copies and $i,j=1,\dots ,d$,
for $i\ne j$.
However, since $\rho_A^{\otimes2}$ is constructed from two identical
copies of $\rho_A$, vectors of the form above cannot be the only contributors
to the RDM. The RDM with the largest possible variance has an equal
weight to all vectors of the form above, which means it will have
the form
\begin{equation}
\rho_A=\otimes_{\alpha\in A}\frac{1}{d}\left(\begin{array}{ccc}
1 & \dots & 1\\
\vdots & \ddots & \vdots\\
1 & \dots & 1
\end{array}\right).\label{eq:wirstRDM}
\end{equation}
Then, the contribution of each site to the first term of the
variance is 
\begin{equation}
\xi_{n}=\left(\left(d+4\binom{d}{2}\right)/d^{2}\right)^{n}.
\end{equation}
The RDM with smallest possible variance will be, for example,
\begin{equation}
\rho_A=\k 0\b 0^{\otimes N_{A}},
\end{equation}
or any other product state in the computational basis. In this case,
the first term of the variance sums up to 1 and $\Var(\hat {p}_{n})=0$.
In both cases A is disentangled from its environment, which demonstrates
that the variance of the estimated value depends
on the basis choice for the vectors $\k v$.

The variance of the flux-resolved moment estimators of the complex valued
random variable defined in 
Eq.~(\ref{eq:fluxEst}) can similarly be computed from
\begin{eqnarray}
|\hat{p}_{n}(\rho_A,\varphi)|^2&=&
\langle V_1 |
e^{i\varphi\hat{Q}_{A}}\rho_A|V_2\rangle
\langle V_2|\rho_A|V_3\rangle\dots 
\langle V_n|\rho_A|V_1\rangle
\nonumber\\
&\times&
\langle V_1| \rho_A| V_n\rangle
\dots \langle V_3|\rho_A| V_2\rangle \langle
V_2 | \rho_A e^{-i\varphi\hat{Q}_{A}}\rho_A |V_1\rangle
\nonumber\\
&=&
 \Tr\left(\left(
 \rho_A e^{i\varphi\hat{Q}_{A}}\otimes 
 e^{- i\varphi\hat{Q}_{A}} \rho_A 
 \mathcal{E}\right)\left(\rho_A^{\otimes2}\mathcal{E})^{n-1}\right)\right), \nonumber\\
\end{eqnarray}
which is bounded from above by the variance for the non-resolved case, as the sum of the first term is composed
of terms with the same absolute values,
but with added phases. The estimator for the variance of the symmetry-resolved
moments is therefore also bounded by 
\begin{eqnarray}
\Var(\hat{p}_{n}(q))&=&\frac{1}{N_{A}}\sum_{\varphi}\Var(\hat{p}_{n}(\varphi))\nonumber\\
&\le& \frac{1}{N_{A}}\sum_{\varphi}\Var(\hat {p}_{n})=\Var(\hat {p}_{n}).\label{eq:cahrgeVar}
\end{eqnarray}

\begin{figure}[t]
\includegraphics[width=.85\linewidth]{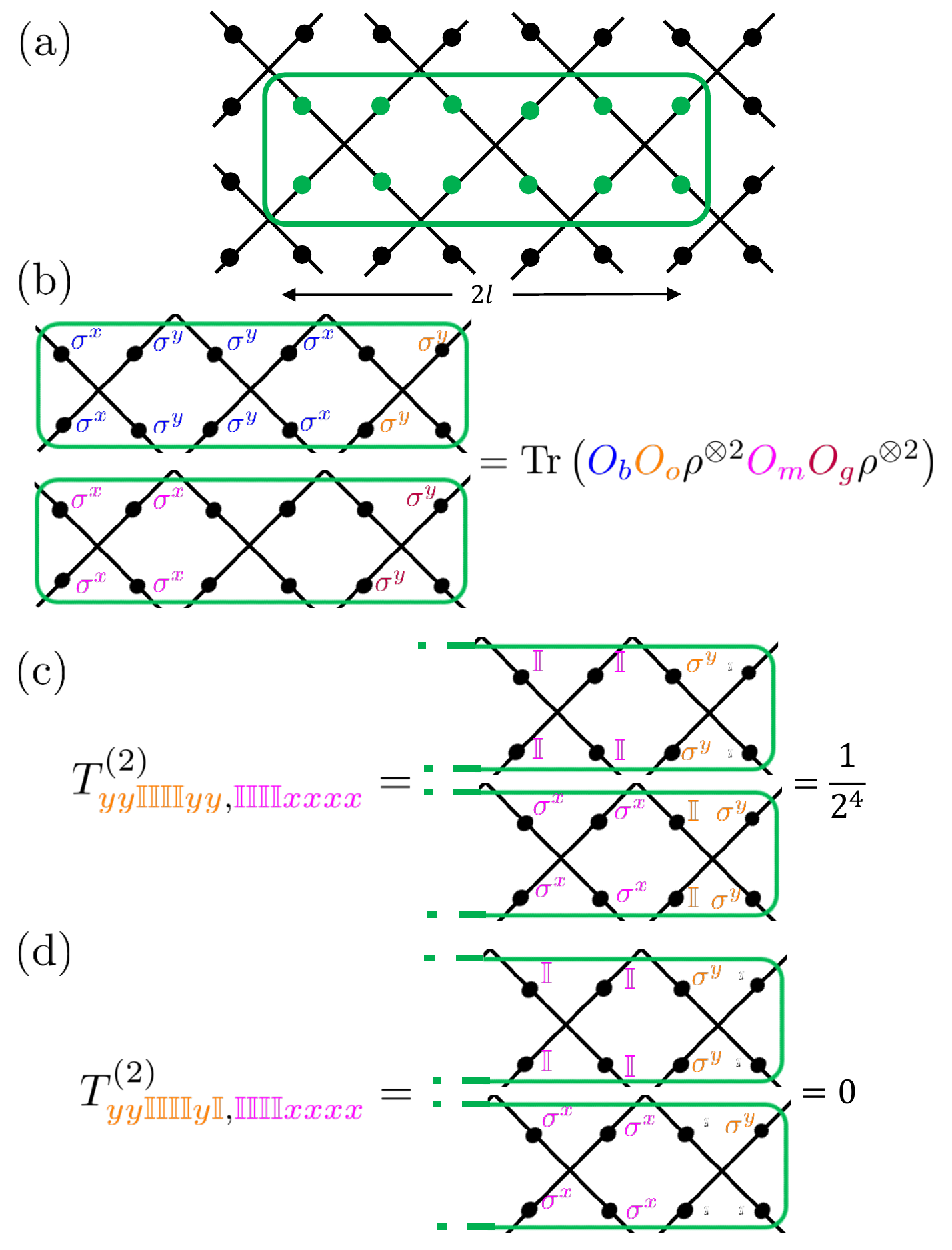}\caption{(a) A configuration of a strip-like subsystem of dimension $2\times2l$, here $l=3$. (b) An example for a contributing configuration of operators to the
term $\protect\Tr((\rho_A^{\otimes2}\mathcal{E})^{n})$
in Eqs.~(\ref{eq:varentropy}) and (\ref{eq:varneg}), for $n=2$ and a
subsystem composed of $l=3$ stars. Depicted in 
pink and blue are stabilizers under which 
the subsystem RDM is invariant, and therefore each of
them can be applied to any of the copies and contribute to the trace
(Note that $\sigma^{y}=i\sigma^{z}\sigma^{x}$). In orange and purple
is a pair of operators which commute with the star operator $\otimes_{i\in s}\sigma_{i}^{x}$
due to having an even number of $\sigma^{y}$ matrices. Such a pair
of operators which commute with the stabilizer operator will also
contribute to the trace, provided they are applied on both sides of the same copy of $\rho_A^{\otimes2}$ factors. 
(c) An example element
of the transfer matrix $T^{(2)}$. Applying the operators in orange
to the $(l+1)$-th star, given that the operator configuration on the $l$-th operator is the one in pink, is allowed, and the expression
will be multiplied by 
$2^{-4}$
due to the four $\frac{1}{2}\sigma^{y}\otimes\sigma^{y}$
applied to the $(l+1)$-th star. 
(d) An additional example element
in the transfer matrix $T^{(2)}$. Applying the operators in orange
to the $(l+1)$-th star will not contribute to the trace, since this
operator configuration do not commute with the plaquette or star terms.}
\label{fig:toricAppendix}
\end{figure}

\section{Explicit variance calculation for the toric code\label{sec:appToricVar}}

Here, we demonstrate how the variance can be calculated exactly in
the toric code model for a subsystem shaped as a narrow strip (Fig. \ref{fig:toricAppendix}a for the partial-basis  distribution, and compare
it to the extracted variance.  We start from Eqs.~(\ref{eq:varentropy}),
(\ref{eq:E21})
and
(\ref{eq:E22}), and the decomposition 
\begin{equation}
\mathcal{E}^{[\alpha]}=\mathbb{I}+\frac{1}{2}\sigma^{x}\otimes\sigma^{x}+\frac{1}{2}\sigma^{y}\otimes\sigma^{y},\label{eq:Edeco}
\end{equation}
which applies for the partial-basis  method in $d=2$. The full $\mathcal{E}$
matrix can be written as 
\[
\mathcal{E}=\sum_{\mathbf{S}}\otimes_{\alpha}S_{\alpha},
\]
where $\mathbf{S}$ is an $N_{A}$-long configuration of the operators
$\mathbb{I},\frac{1}{2}\sigma^{x}\otimes\sigma^{x},\frac{1}{2}\sigma^{y}\otimes\sigma^{y}$. 
We use the local symmetry of the toric code ground state
\begin{equation}
\otimes_{i\in s}\sigma_{i}^{x}\rho=\otimes_{i\in p}\sigma_{i}^{z}\rho=\rho\otimes_{i\in s}\sigma_{i}^{x}=\rho\otimes_{i\in p}\sigma_{i}^{z}=\rho,\label{eq:toricsym}
\end{equation}
in order to distinguish configurations $\mathbf{S}$ which will contribute
to Eq.\ (\ref{eq:varentropy}). An allowed configuration $\mathbf{S}$
will contain any number of star ($\otimes_{i\in s}\sigma_{i}^{x}$)
or plaquette ($\otimes_{i\in p}\sigma_{i}^{z}$) operators, also called
the stabilizers. Pairs of some operator which commutes with the stabilizers
and act on two sides of the same copy of $\rho^{\otimes2}$ are also
allowed, as well as combinations of operators which can be transformed
into such pairs by a multiplication of the operators by stabilizers.
This is illustrated in Fig.~\ref{fig:toricAppendix}b.

\begin{figure}[!tp]
    \centering
    \includegraphics[width=0.9\linewidth]{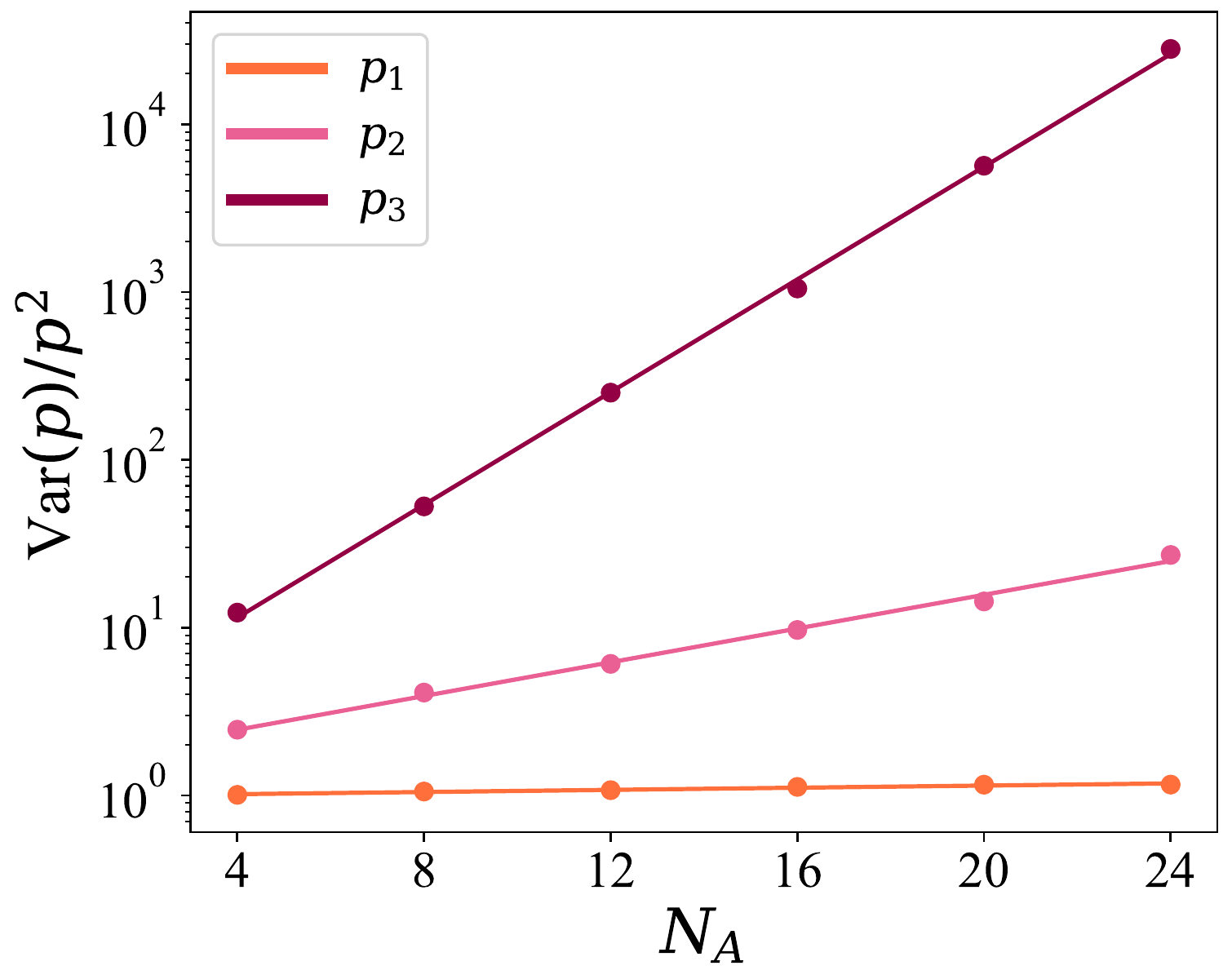}
    \caption{The estimated variances of a strip-like system in the toric code model.}
    \label{fig:vars_appendix}
\end{figure}

\begin{table}[!tp]
\begin{tabular}{|c|c|c|}
\hline 
 & Exact variance & Estimated variance\tabularnewline
\hline 
\hline 
$n=1$ & $O\left(1.016^{N_{A}}\right)$ & $O\left(1.016^{N_{A}}\right)$\tabularnewline
\hline 
$n=2$ & $O\left(1.225^{N_{A}}\right)$ & $O\left(1.220^{N_{A}}\right)$\tabularnewline
\hline 
$n=3$ & $O(1.549^{N_{A}})$ & $O\left(1.564^{N_{A}}\right)$\tabularnewline
\hline 
\end{tabular}\caption{Exact dependence of the variance on system size for the toric code
model as calculated in by the transfer matrix method compared to the
variance estimated from sampling of the expressions in Eqs.~(\ref{eq:varentropy}) and (\ref{eq:varneg}).}

\label{table:toric}
\end{table}

We calculate the variance for a narrow system of dimension $2\times 2l$, as depicted in Fig.~\ref{fig:toricAppendix}a.
Such a system is composed of a chain of $l$
contiguous stars. For a moment of rank $n$, we think of $n$ copies
of the subsystem, and write a $3^{4n}$-dimensional vector of combinations
of operators $\left[\mathbb{I},\sigma^{x}\otimes\sigma^{x},\sigma^{y}\otimes\sigma^{y}\right]$
on a single star in all $n$ copies. We now write a transfer matrix
$T^{(n)}$ which takes the operator combinations on the $l$-th star
to the contributing combinations on the $(l+1)$-th star: $T_{\mathbf{i},\mathbf{j}}^{(n)}$
equals the contribution of an operator combination with operators
$\mathbf{j}$ on the $(l+1)$-th star to the variance in Eq.\ (\ref{eq:varentropy}),
given that the combination on the $l$-th star is $\mathbf{i}$, where
the symmetries in Eq.\ (\ref{eq:toricsym}) are considered, as well
as the $\frac{1}{2}$ factors in Eq.\ (\ref{eq:Edeco}). For clarity,
we give a specific example in Fig.~\ref{fig:toricAppendix}c-d. With
these definitions, the first term in the left side of Eq.\ (\ref{eq:varentropy})
for a subsystem of $2\times2l$ sites is $\b{c_{0}}
(T^{(n)})^{l-1}\k{c_{0}}$,
where $\k{c_{0}}$ is the vector of allowed contributions for the
edges of the subsystem as can be deduced from the ground states of
the toric code model in Eq.\ (\ref{eq:gs}). The dependence of this
term in the system size is thus $O\left(\lambda_{\max}^{4l}\right)$,
where
\begin{equation}	
	\lambda_{\rm max}:= \| T^{(n)}\|
\end{equation}
is the largest eigenvalue of the Hermitian	
$T^{(n)}$. 
One may, in fact, work with equivalent but much smaller transfer matrices,
by considering only the two left sites of a star rather than the whole
star. This allows decreasing the transfer matrix to dimension $3^{2n}\times3^{2n}$.
For $n\le3$ the matrix $T^{(n)}$ can be extracted and diagonalized
exactly. 
We have performed an estimation of the variance for such strip-like systems, similarly to the one done in Fig.~\ref{fig:variance} in the main text. The estimated variances are displayed in Fig.~\ref{fig:vars_appendix}.
We have compared the results obtained exactly using the transfer
matrix to the numerical variances and
got a good agreement, demonstrating the accuracy of our PEPS calculations,
as can be seen in the Table \ref{table:toric}.

\bibliography{many-je}

\end{document}